\documentclass[prb,twocolumn,aps,amsmath,amssymb]{revtex4}
\usepackage{graphicx}
\usepackage{bm}

\begin{document}


\title{Transport properties of diluted magnetic semiconductors:
  Dynamical mean field theory and Boltzmann theory}

\author{E. H. Hwang and  S. Das Sarma}

\affiliation{Condensed Matter Theory Center,
Department of Physics, University of Maryland, College Park, 
MD 20742}

%
\begin{abstract}

The transport properties of diluted magnetic semiconductors (DMS)  
are calculated using dynamical mean field theory (DMFT) and Boltzmann
transport theory.
Within DMFT we study the density of states and the dc-resistivity,
which are strongly parameter dependent such as temperature, 
doping, density of the carriers, and 
the strength of the carrier-local impurity spin exchange coupling.
Characteristic qualitative
features are found distinguishing weak, intermediate, and 
strong carrier-spin coupling and allowing quantitative determination of 
important parameters defining the underlying ferromagnetic mechanism. 
We find that spin-disorder scattering, formation of bound
state, and the population of the minority spin band are all
operational in DMFT in different 
parameter range. We also develop a complementary  Boltzmann
transport theory for scattering by screened ionized impurities. The
difference in
the screening properties between paramagnetic ($T>T_c$) and
ferromagnetic ($T<T_c$) states gives rise to the 
temperature dependence (increase or
decrease) of resistivity, depending on the carrier density, as
the system goes from the paramagnetic phase to the ferromagnetic phase.
The metallic behavior below $T_c$ for optimally doped DMS samples can
be explained in the Boltzmann theory by temperature dependent
screening and thermal change of carrier spin polarization.



\end{abstract}
\vspace{.25cm}

\maketitle

\section{introduction}

Diluted magnetic semiconductors 
(DMS) \cite{Ohno98,Ohno96,Matsukura98,Reed01}
with transition temperatures as high as 150K
(GaMnAs at $5\%$ Mn \cite{Ohno98,Ohno96,Matsukura98}) 
or even above room temperature (GaMnN, GaMnP\cite{Reed01}) 
are attracting much attention lately 
in part because of possible `spintronic' applications \cite{dassarma01}. 
The prototypical DMS material is Ga$_{1-x}$Mn$_x$As with the Mn ions
substitutionally replacing Ga at the cation sites.
It is widely believed that the ferromagnetism in this material is
carrier induced with 
holes donated by Mn ions mediating a ferromagnetic interaction between
the randomly localized Mn spins \cite{dassarma03}.
The coupling of carrier spins (holes in GaMnAs) and localized
moments (Mn impurities) 
gives rise to the unique magnetic and transport (as well as optical)
properties in DMS.

DMS transport properties are influenced by the exchange
interaction between the carriers and the localized moments as spin
fluctuation scattering contributes to the resistivity.
The experimentally measured dc-resistivity in DMS materials
\cite{Matsukura98,oiwa,hirakawa} shows 
interesting behavior strongly depending on the concentration of the
magnetic impurity and temperature.
In  In$_{1-x}$Mn$_x$As \cite{hirakawa} and
Ga$_{1-x}$Mn$_x$As \cite{Matsukura98,oiwa} with low Mn concentration
($x<0.03$)  only an insulating behavior has been observed in transport
measurements. 
However, near  optimal doping $x \approx 0.05$, where the highest
value of $T_c$ is reported, the non-monotonic behavior
(insulator-metal-insulator) of the
resistivity as a function of temperature is observed. 
A resistivity peak appears near the critical temperature ($T_c$), and
the resistivity shows metallic behavior ($d\rho_{dc}/dT >0$) below
$T_c$ and 
insulating behavior ($d\rho_{dc}/dT <0$) at higher temperatures.
The peak has been understood as the critical scattering effects at
$T_c$ of spin fluctuations \cite{deGennes}. 

In this paper we present theoretical calculations 
for DMS transport properties  and study the role of the
carrier-spin coupling 
which are crucial for ferromagnetic properties.
Transport measurements have proven useful in 
understanding the physics of the colossal magnetoresistance (CMR) 
manganites
where carrier-spin coupling is also crucial \cite{CMRReview}. 
In our calculation we use a recently developed non-perturbative method,
the ``dynamical mean-field theory'' (DMFT),\cite{Kotliar}
for calculating the DMS transport properties.
A non-perturbative method is needed in DMS materials because the
crucial physics involves  
bound-state formation and other aspects of intermediate carrier-spin 
couplings not accessible to perturbative methods.
(We note that the most interesting phenomena in DMS involve 
intermediate couplings and intermediate temperatures.
This regime is very difficult to treat by 
standard analytical or numerical methods.)
The DMFT has been recently applied to the
DMS system to calculate the magnetic transition temperature and the
optical conductivity \cite{Chattopadhyay01a,hwang}.  DMFT is
essentially a lattice 
quantum version of the Weiss mean field theory where the appropriate
density of states (including impurity band formation) along with
temporal fluctuations are incorporated within an effective local field
theory.
An important ingredient of DMFT \cite{Chattopadhyay01a} 
is that it reduces to the
standard RKKY physics in the weak-coupling regime and the double
exchange physics in the strong coupling regime.

Our DMFT results show interesting  dependence 
of resistivity on carrier-spin coupling ($J$), carrier density ($n$), 
doping of the magnetic material ($x$), 
and temperature ($T$) revealing key features of the underlying physics.
We find that our results show
many similarities to colossal magnetoresistance (CMR) 
manganites \cite{CMRReview}, especially,  
for a large carrier-spin coupling and near half filling of the
impurity band (i.e. in the double exchange regime)
since the spin-disorder scattering dominates in this
parameter range. However, the DMS dc-resistivity
exhibits novel features not found in the CMR in the weak
coupling RKKY limit ($J \le 1.0t$, where $t$ is the band width of the
carrier). These novel features arise mainly from the 
bound state formation of the carrier-local spins and carrier
occupation of the minority spin band. The formation of the bound state
gives rise to an insulating behavior even
if the carrier localization effects are not taken into account.
Experimental observation of our predictions should lead to crucial 
information about bound state formation and impurity band physics in 
this problem.
Even though we focus on III-V compound based DMS such as
Ga$_{1-x}$Mn$_x$As,
the results presented in this paper are general for all DMS materials.

An important limitation of DMFT is that it is a non-perturbative {\it
local} theory that cannot really incorporate {\it spatial}
fluctuations, and therefore resistive charged impurity scattering
with its strong momentum dependence is essentially impossible to
handle in DMFT.
We therefore consider the Boltzmann transport theory to calculate the
resistivity of the DMS systems \cite{brey,brey2}. 
The Boltzmann theory is used to calculate the charged impurity
disorder limited DMS resistivity (whereas the DMFT is used for
obtaining the spin disorder limited DMS resistivity). Charged impurity
disorder arises here both from the ionized Mn acceptor in GaMnAs as
well as other charged defects/impurities invariably present in a
semiconductor. 
Using relaxation time approximation
we assume that the Boltzmann resistivity is due to ionized impurities.
The carriers (holes) are scattered by the screened Coulomb
potential, which we calculate using the linearized
Thomas-Fermi (TF) approximation and the random phase approximation
(RPA). We find that the dominant 
temperature dependence of the resistivity comes from
the change in the screening length in the high density metallic samples. 
In the metallic GaMnAs DMS samples the change in the measured 
resistivity, when
the temperature goes from $T_c$ to zero, is about 20\% in good
agreement with our calculation. 
In the ferromagnetic state ($T<T_c$), as the temperature decreases from
$T=T_c$, we find strong temperature dependence of the resistivity
arising from the low temperature screening function and the 
thermal change of the carrier
densities in each spin-split subband.  
When all the carriers are polarized (this happens in the low
density limit) the screening function
is almost independent of temperature and the resistivity is also  
temperature independent.
Inclusion of both spin disorder and ionized impurity disorder in DMS
transport is the important ingredient of our theory.

This paper is organized as follows. In Sec. II, we describe our model
and our theoretical approach based on the dynamical mean-field theory
(DMFT). In Sec. III, we study, in detail, the density of states and
describe the formation of the spin polarized impurity band. In
Sec. IV, we provide the results of calculated DMS resistivity within
DMFT. In Sec. V we calculate the transport resistivity within the
Boltzmann transport theory  for ionized impurities.
In Sec. VI, we summarize our qualitative findings and
providing a critical discussion of the applicability of our results to
DMS systems. A brief conclusion is given in Sec. VII.

\section{Model and formalism}

Our basic model of DMS systems is that of 
magnetic dopants (``impurities'') interacting through a local exchange
coupling with carriers in the host
semiconductor material. 
The generally accepted Hamiltonian of the system is given by
\begin{equation}
H  =  H_{host}  + H_M + H_{AF},
\label{ham}
\end{equation}
where $H_{host}$ describes carrier propagation in the host 
semiconductor band.
For simplicity we consider a host material with a single 
non-degenerate band. We therefore write
\begin{equation}
H_{host}= \sum_{\alpha}
\int d^3x \psi _{\alpha }^{+}(x) \left [ \frac{\nabla ^{2}}{2m}
+ V_R(x) \right ] \psi _{\alpha }(x),
\end{equation}
where $\alpha$ is the spin index and $V_R$ is a random potential
arising from non-magnetic defects  in the material (e.g., As antisite
defects, unintentional background charged impurities, etc.). 
The second (magnetic) 
term in Eq. (\ref{ham}), $H_M$, describes coupling of the carriers to 
an array of impurity (e.g. Mn) spins at positions $R_i$,
\begin{equation}
H_M = \sum_{i,\alpha,\beta}
\psi _{\alpha }^{\dag}(R_i)
\left [ J\hat{\bf S}_{i}\cdot {\bm \sigma}_{\alpha \beta } + W
  \delta_{\alpha,\beta} \right ]
\psi_{\beta }(R_i),
\end{equation}
where $J$ is the local exchange coupling between the spin of the
magnetic impurity and the the spins of the semiconductor carriers,
$W$ is the (Coulombic) potential arising from the magnetic dopant,
$R_i$ are the positions of the magnetic dopants, and $\sigma$ is the
Pauli matrix.
Here we absorb the magnitude of the 
impurity spin into the coupling $J$ (which we take to be positive), 
and represent the spin 
direction by the unit vector $\hat{\bf S}$. 
The third term in Eq. (\ref{ham}), $H_{AF}$, is
the direct Mn$-$Mn short-range antiferromagnetic exchange
interaction 
\begin{equation}
H_{AF} = \sum_{i,j}J_{AF}(R_i-R_j){\bf S}_i {\bf \cdot} {\bf S}_j,
\end{equation}
where $J_{AF}$ is a direct antiferromagnetic exchange coupling between
impurity spins. 

In this section we approximate our model by neglecting the nonmagnetic
random potential, $V_R$, and  the direct antiferromagnetic exchange
interaction, $J_{AF}$.
Lattice defects may be playing an important role in determining
magnetic and transport properties of the samples, but we assume 
here that these defects enter our theory only in determining the basic
parameters of the model, namely, the density of magnetically active
dopants $n_{\textrm{i}}$, the hole density $n_{\textrm{c}}$, and
perhaps the local effective exchange coupling $J$ between the holes
and the magnetic impurities, and do not include any defects into our
model explicitly.  We include charged impurity scattering through the
Boltzmann equation in section IV of this paper.
We believe that the effects of the antiferromagnetic coupling between
magnetic impurities are either negligibly small or
incorporated into the effective parameters of the model.  
Actually, in
the parameter range of interest to us ($x \ll 1$), where DMS
ferromagnetism typically occurs, the magnetic impurities are separated
from each other by non-magnetic atoms, and this 
short-range antiferromagnetic
interaction, which rapidly decays with the distance, should be
negligible.  
These approximations are nonessential and are done in
the spirit of identifying the minimal DMS magnetic model of interest.
Both of these effects, which may be of quantitative importance in some
situations, can be included in the theory by adjusting the parameters
of the model or perhaps at the cost of introducing more unknown
parameters characterizing these interactions.  Recently
several theories
for DMS ferromagnetism explicitly including spatial disorder effects
and antiferromagnetic coupling have been developed \cite{priour}.

In our DMFT DMS 
model there are two sources of coupling between the carrier and the
impurity magnetic moment: a spin-spin coupling ($J$)
and a potential  scattering ($W$).
The crucial physical issues are revealed by the consideration of a 
ferromagnetic state in which all impurity spins $S_i$ are aligned, 
say, in the $z$ direction. Then the carriers with spin parallel to 
$S_i$ feel a potential $-J+W$ on each magnetic impurity site and
anti-parallel carriers feel a potential $J+W$. These potentials 
self-consistently
rearrange the electronic structure. The spin-dependent part of this 
rearrangement provides the energy gain which stabilizes the 
ferromagnetic state. The key physics issue is, evidently, whether the 
potential $W\mp J$ is weak (so its effect on carriers near the lower 
band edge is simply a scattering phase shift) 
or strong (so only majority spin or perhaps both 
species of carriers are confined into spin-polarized impurity bands). 
Recent density functional supercell calculations \cite{Sanvito01}
suggest that in GaMnAs
$-J+W$ is close to the critical value for bound state formation
for the majority spin systems.
Unfortunately, the precise effective values of $J$ and $W$ are
typically unknown in a DMS system, and may have to be extracted
experimentally.

We assume that magnetic impurities under consideration enter
substitutionally at the cation sites (\emph{e.~g.} Mn impurities at Ga
sites) and the III$_{1-x}$Mn$_x$V system as a lattice of sites, which
are randomly nonmagnetic (with probability $1-x$) or magnetic (with
probability $x$), where $x$ now indicates the relative concentration
({\it i.e.} per Ga site) of active Mn local moments in IIIMnV. 
If more complete information about Mn locations on the GaAs lattice
becomes available it will be straightforward to incorporate that in
the DMFT formalism.

We now introduce the DMFT for the Hamiltonian given by Eq. (\ref{ham}).
Within the general scheme of the DMFT, the local (momentum
independent) self energy  of the system, 
$\Sigma(i\omega_n)$,
can be obtained from the time dependent mean field function.
Then the single particle Green function is approximated by
\begin{equation}
G({\bf k},i\omega_n) = \frac{1}{i\omega_n - (\epsilon_{\bf k}-\mu) -
  \Sigma(i\omega_n)},
\label{green}
\end{equation}
where $\mu$ is the chemical potential.
With the local self energy all of the relevant
physics may  be determined from the local (momentum-integrated) Green
function defined by
\begin{eqnarray}
G_{\textrm{loc}}(i\omega_n) & = & \int \frac{d^3k}{(2\pi)^3} G({\bf
  k},i\omega_n) \nonumber \\
& =& \int d\epsilon D(\epsilon) \frac{1}{
i\omega_n +\mu  - \epsilon 
-\Sigma_{\sigma}(i\omega_n)},
\label{G_loc}
\end{eqnarray}
where $D(\epsilon)=\int d^dp/(2\pi)^d \delta (\epsilon-\epsilon_p)$ is
the density of states (DOS) for the noninteracting 
system. The information of the lattice geometry is included through
the noninteracting DOS. 

In our model $G_{\textrm{loc}}$ is a matrix in spin
index and depends on whether one is considering a magnetic
($a)$ or non-magnetic ($b)$ site. Since $G_{\textrm{loc}}$ is a local function,
it is the solution of a local problem specified by a mean-field
function $g_0$, which is related to the partition function
$Z_{\textrm{loc}}=\int d \hat{\textbf{S}} \exp(-S_{\textrm{loc}})$ with action
\begin{eqnarray}
S_{\textrm{loc}} & = & \int d\tau \int d\tau'
\sum_{\alpha \beta }c_{\alpha }^{+}(\tau )
\left [ g_{0\alpha \beta }^{a}(\tau-\tau ^{\prime }) \right ] 
c_{\beta }(\tau ^{\prime }) \nonumber \\
& + & \int d\tau \sum_{\alpha,\beta} c_{\alpha }^{+}(\tau )
\left [ J \hat {\textbf{S}} \cdot \bm{\sigma}_{\alpha  \beta} 
+ W \delta_{\alpha,\beta} \right ] c_{\beta }(\tau)
\end{eqnarray}
on the $a$ (magnetic) site and 
\begin{equation}
S_{\textrm{loc}}=\int d\tau \int d\tau'
\sum_{\alpha \beta } g_{0\alpha \beta }^{b}(\tau
-\tau ^{\prime }) 
c_{\alpha }^{+}(\tau)c_{\beta }(\tau^{\prime}),
\end{equation}
on the non-magnetic ($b$) site.
Here $c_{\alpha}(\tau)$
($c_{\alpha}^{+}(\tau)$) is the destruction (creation) operator of a
fermion in the spin state $\alpha$ and at time $\tau$.
$g_0(\tau - \tau')$ plays the role of the Weiss mean field (bare
Green function for the local effective action $S_{\textrm{loc}}$) and is
a function of time.
$G_{loc}$ depends only on frequency and is therefore the solution of a
single-site problem. The local Green function $G_{loc}$ of the
effective single-site problem is solely determined by the partition
function, $Z_{\textrm{loc}}$, namely,
$G_{loc}(i \omega_n) = \delta \ln Z_{\textrm{loc}}/{\delta
g_{0}^{a}}$ 
which is identical to the
local Green function computed by performing the momentum integral
using the same self energy. Then the self-energy is defined by
\begin{equation}
\Sigma_{\alpha \beta} (i\omega_n) = g_{0\alpha \beta} (i\omega_n) -
G_{loc, \alpha \beta}^{-1}(i \omega_n).
\label{self}
\end{equation}

The $a$-site mean-field function $g_{0}^{a}$ can be written as
$g_{0\alpha \beta }^{a}=a_{0}+a_{1}\hat {\textbf{m}}\cdot
\bm{\sigma}_{\alpha \beta }$ with $\hat {\textbf{m}}$ the
magnetization direction 
and $a_{1}$ vanishing in the paramagnetic state ($T>T_c$). 
Since the spin axis is chosen parallel to $\hat {\bf m}$ $g_0^a$ becomes a
diagonal matrix with components parallel ($g_{0\uparrow}^a = a_0 +
a_1$) and antiparallel ($g_{0\downarrow}^a = a_0 - a_1$) to $\hat {\bf m}$.

The form of the dispersion given in full Hamiltonian
Eq.~(1) applies only near the band edges.  It is
necessary for the method to impose a momentum cutoff, arising
physically from the carrier band-width.  We impose the cutoff by
assuming a semicircular density of states $D(\epsilon)=a_0^{3}\int
\frac{d^{3}p} {\left( 2\pi \right) ^{3}}\delta (\varepsilon
-\varepsilon _{pa}) = \sqrt{4t^2-\epsilon^2}/2\pi t$ with
$t=(2\pi)^{2/3}/ma_0^2$.  The parameter $t$ is chosen to correctly
reproduce the band edge density of states.  Other choices of upper
cutoff would lead to numerically similar results.  This choice of
cutoff corresponds to a Bethe lattice in infinite dimensions.
Other (perhaps more realistic) choices for the density of states would
give results qualitatively similar to our results since
the band edge density of states has the correct physical behavior in
our model. For this
$D(\epsilon)$ the self consistent equation for $g_{0}$ obeys the
equation
\begin{eqnarray}
{\bf g}_{0}^{a}(i\omega_n)&=&i\omega_n +\mu
-(1-x) \left \langle {\bf g}_{0}^{b}(i\omega_n)^{-1} \right \rangle
\nonumber \\ 
& - & x \left \langle \left[ {\bf g}_{0}^{a}(i\omega_n)
+ \left ( J\hat{\bf S}\cdot {\bm \sigma}
_{\alpha \beta }+ W \right )  \right]^{-1} \right \rangle,
\end{eqnarray}
where the angular brackets denote averages performed in the ensemble
defined by the appropriate $Z_{\textrm{loc}}$, i.e.,
$\langle A \rangle = \int d \hat{\bf S}  P(\Omega) A$,
with 
$P(\Omega) = \exp(-S_{\textrm{loc}})/Z_{\textrm{loc}}$.
With these mean field functions the local Green function can be
written as
\begin{equation}
G_{\textrm{loc}}(i\omega_n)=i\omega_n + \mu - {\bf g}_0.
\end{equation}
The self energies are evaluated using Eq. (\ref{self}) and the full
Green function from Eq. (\ref{green}).  Physical observables can be
obtained from the full Green function $G({\bf k},\omega)$. In particular, 
the mean field function $g_{0\sigma}$ can be easily
calculated at $T=0$
\begin{equation}
g_{0\sigma}(\omega) = \omega +\mu
 -  x \frac{1}{g_{0\sigma}-(W \mp J)} - (1-x)\frac{1}{g_{0\sigma}},
\end{equation}
and $T \ge T_c$
\begin{equation}
g_{0\sigma}(\omega) = \omega +\mu
 -  x \frac{g_{0\sigma}-W}{(g_{0\sigma}-W)^2 - J^2} - 
(1-x)\frac{1}{g_{0\sigma}}.
\end{equation}


\section{density of states}

The density of states plays crucial roles in determining the physical
properties of the DMS system. Especially, the formation of the
impurity band arising from the impurity doping 
gives rise to many different aspects from the continuum (i.e. virtual
crystal approximation) semiconductor band model. 
In this section, we calculate the DOS for different parameters ($J$,
$x$, $W$, and $T$) and show how the impurity bands are formed and separated
from the main band.
We describe the DOS of dynamical mean-field calculations applying to simple
semi-circle models.  
The DOS is given by the imaginary part of the Green function
\begin{equation}
D_{\sigma}(\omega) = -\frac{1}{\pi} {\mathrm {Im}} G_{\sigma}(\omega).
\end{equation}

In our model for a strong magnetic coupling $J > J_c$ two spin
polarized impurity bands appear at the bottom (majority spin) and at
the top (minority spin) of the main band.  Each isolated impurity band
has the 
weight $x$. However, if the coupling is not strong ($J\le J_c$)
the impurity band is not completely separated from the main bend.
All DMS samples show that the carrier density is much smaller
than the impurity concentration ($n < x$) due to the heavy
compensation. As the system is the partially
compensated, the chemical potential $\mu$ is located in the lower
impurity band (if the impurity bands are formed) or the lower band
edge (if the bands are not formed). Thus all physical properties are
determined in the lower energy band edge. Throughout this paper we
only show the DOS near the lower energy band.

\begin{figure}
\includegraphics[width=2.6in]{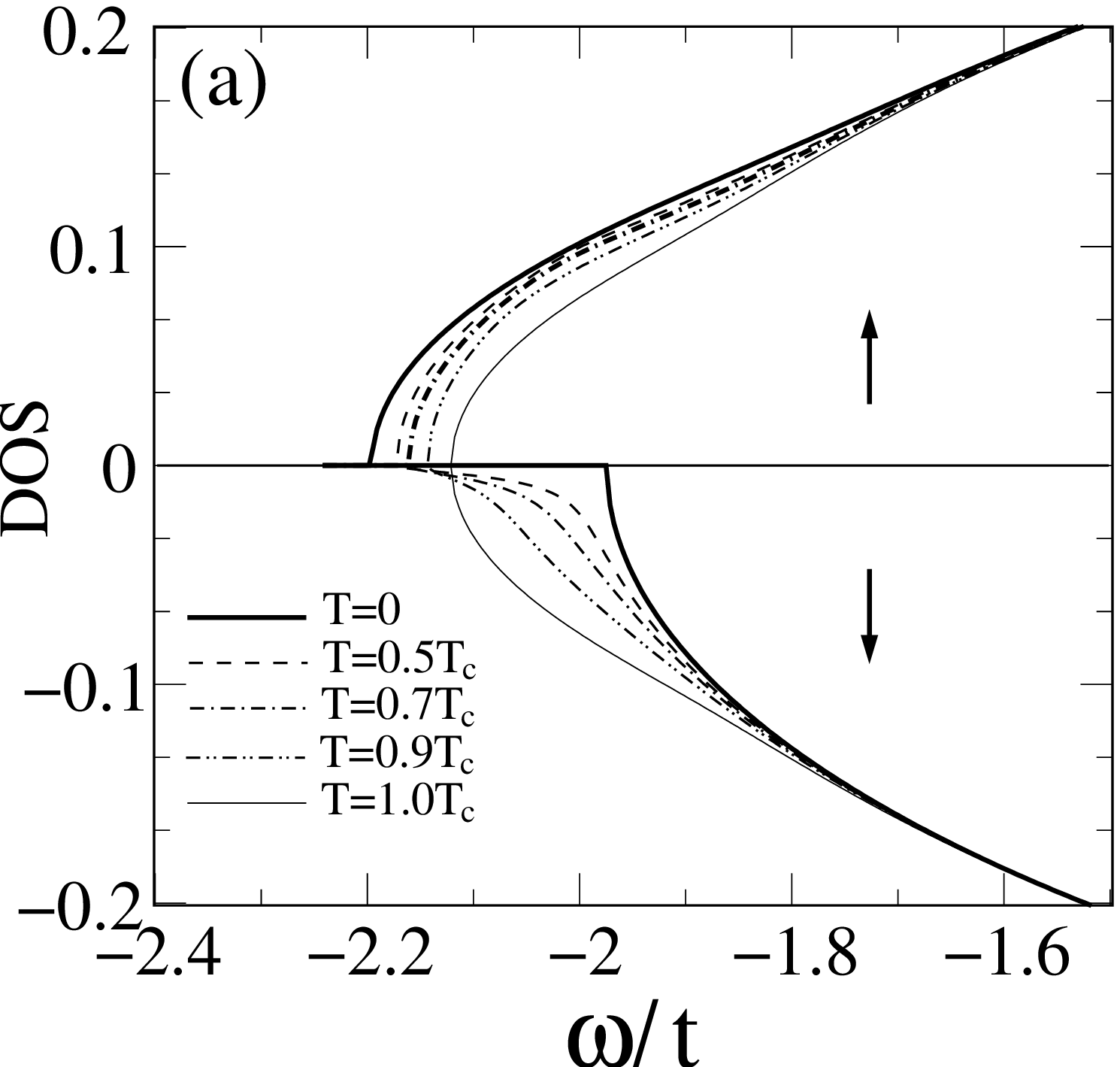}
\includegraphics[width=2.6in]{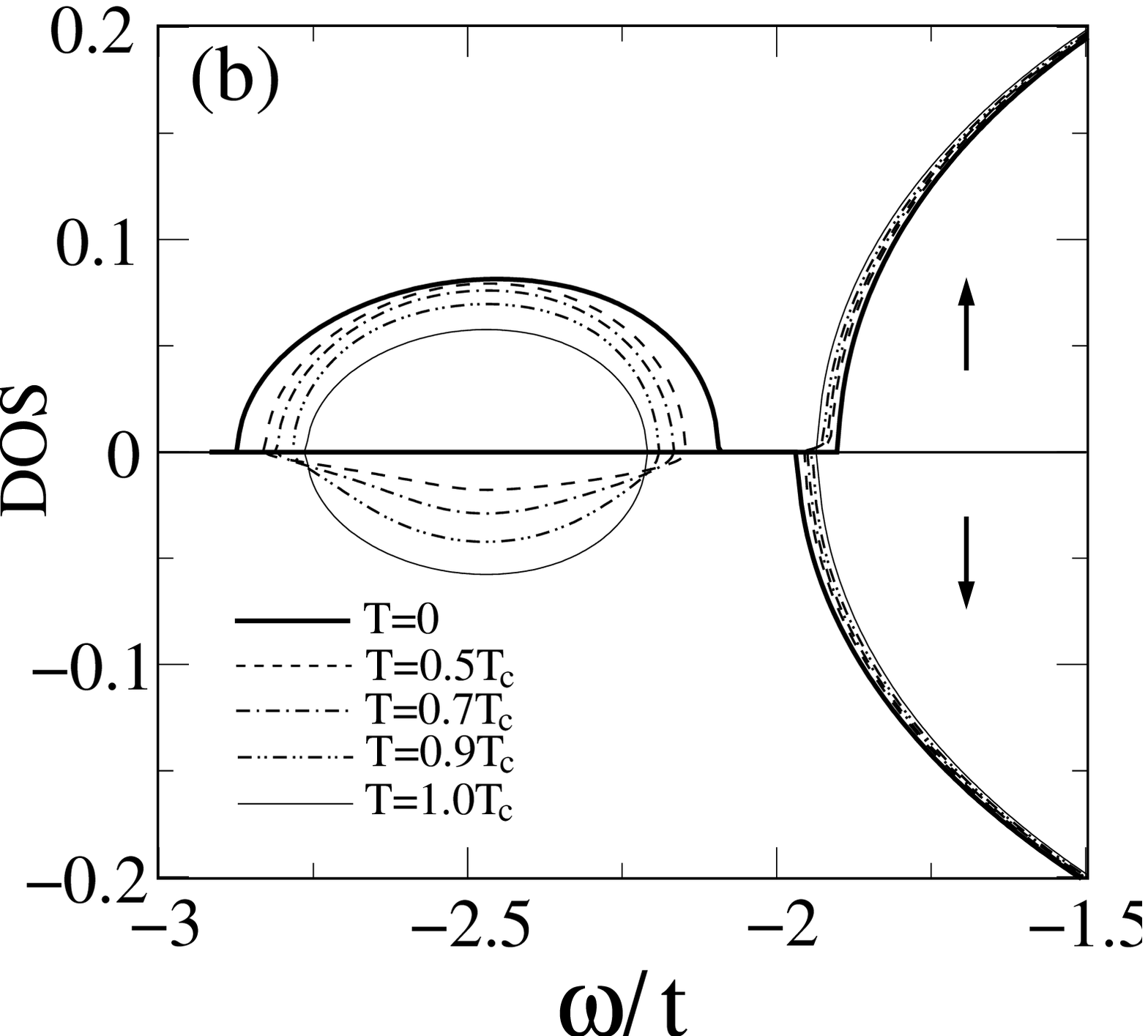}
\caption{
\label{fig_1}
Temperature dependence of the DOS for $x=0.05$ and 
for a fixed coupling (a) $J=1.0t$ and (b) $J=2.0t$.
The evolutions of majority (minority) spin DOS show in top (bottom) panels
for various temperature $T/T_c=$0.0, 0.5, 0.7, 0.9, and 1.0.
}
\end{figure}


\begin{figure}
\includegraphics[width=2.6in]{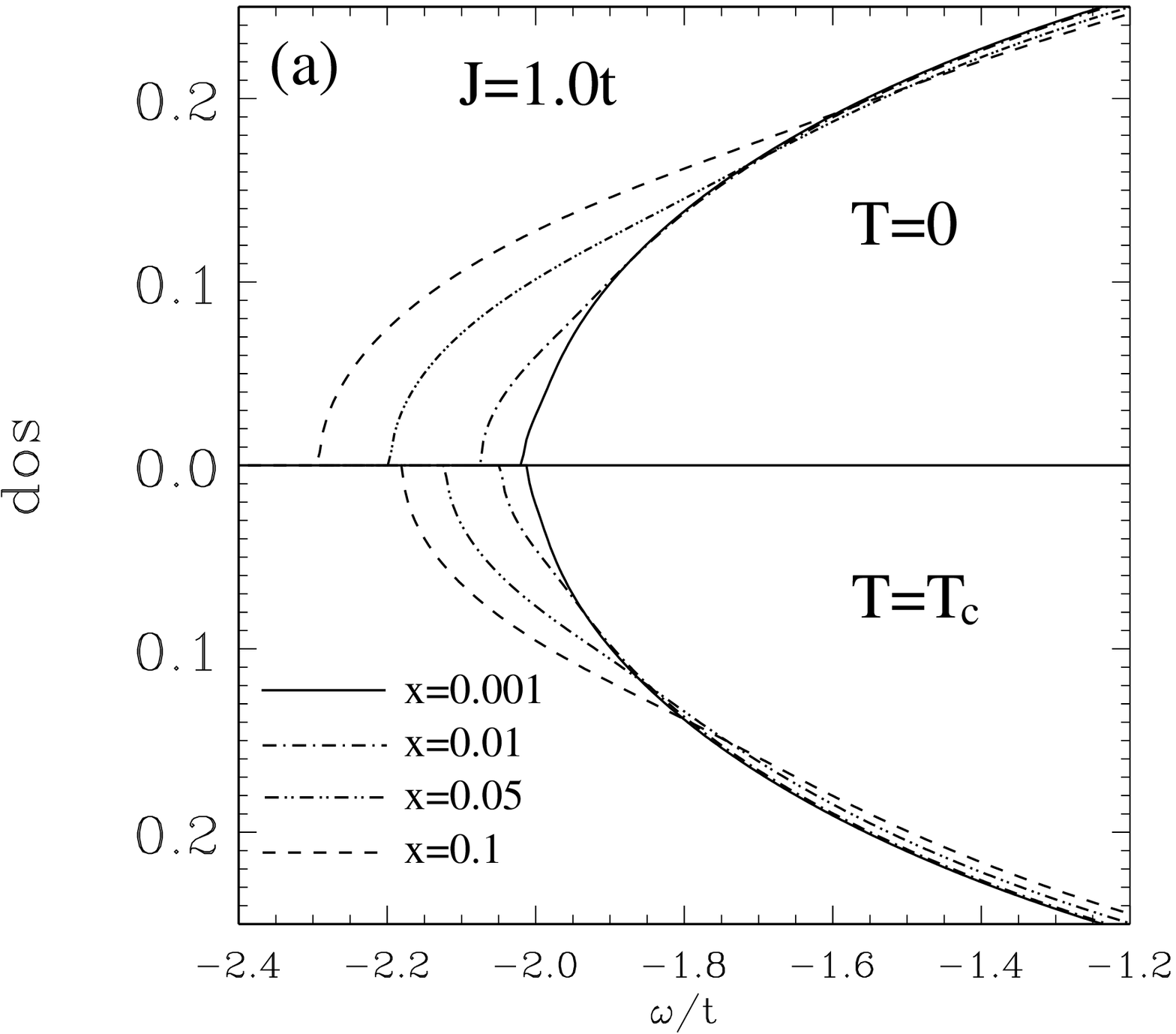}
\includegraphics[width=2.6in]{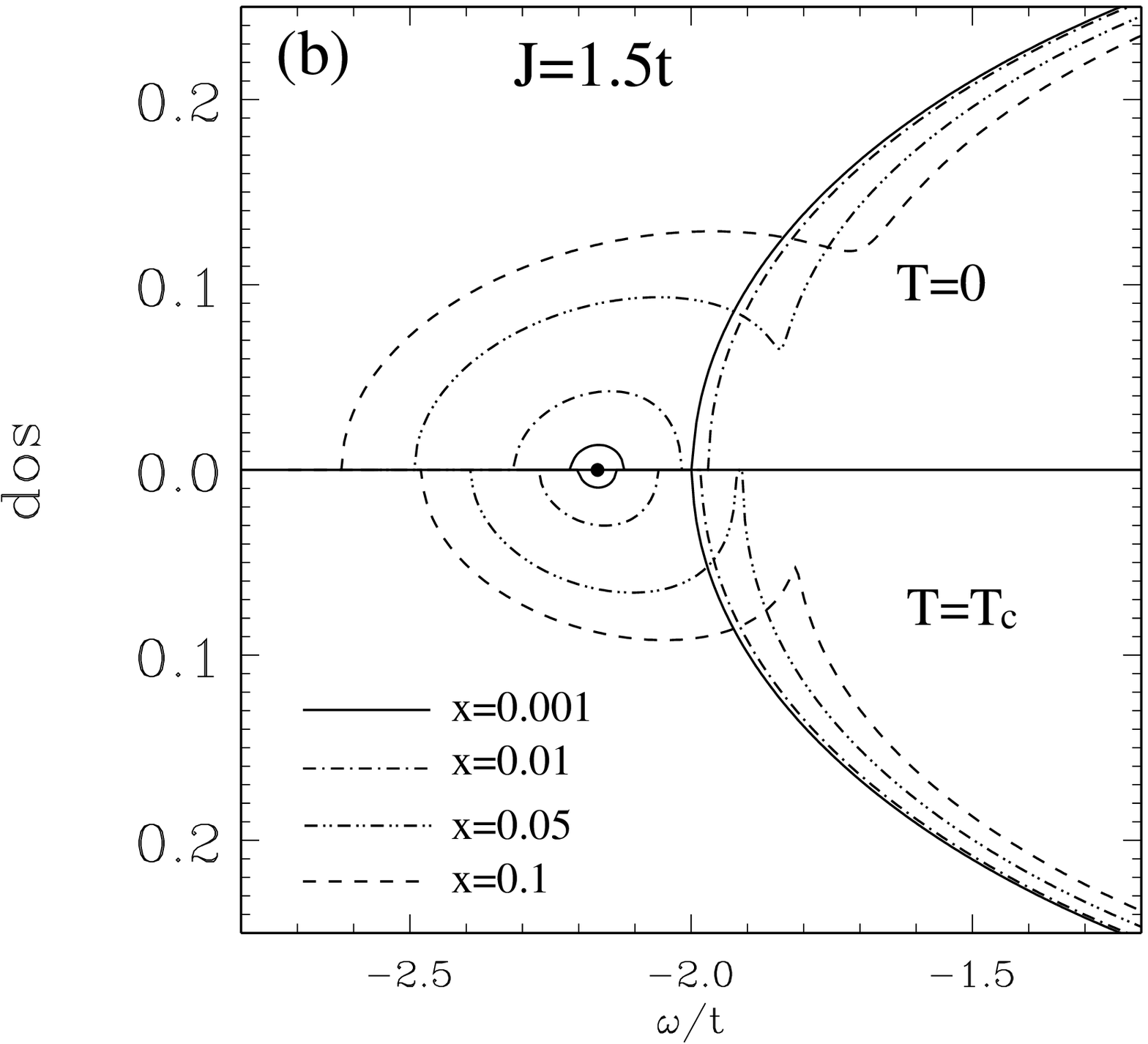}
\caption{
\label{fig_2}
The calculated majority spin DOS at $T=0$ (upper half) and $T=T_c$
(lower half) is shown for various doping $x=$ 0.001, 0.01, 0.05, 0.1
and a fixed coupling (a) $J=1.0t$, and (b) $J=1.5t$. 
}
\end{figure}


In Fig. \ref{fig_1} we show the calculated DOS for various
temperatures as a function of energy.
The evolutions of majority (minority) spin DOS are shown 
in top (bottom) panels. 
In Fig. \ref{fig_1}(a) the strength of the coupling constant ($J=1.0t$)
is not strong enough to form the impurity band. Note that this value of
coupling constant ($J=1.0t$) is the critical value for impurity band
formation as $x \rightarrow 0$. At $T=0$ the majority (minority) spin band 
is shifted to lower (higher) energy compared with the noninteracting
band which has 
a band edge at $\omega = -2.0t$. Thus all carriers are fully polarized
when the carrier density is low. As the carrier density increases they
start to occupy the minority band if the chemical potential 
crosses into the minority spin band. Recently we showed that
the optical conductivity of the system is dramatically changed with
the occupation of the minority band \cite{hwang}. We show in this
paper that the calculated dc transport properties are also very
sensitive to the minority band occupation.  
As the temperature increases, the minority band occupation grows 
and the carriers with minority spin increase due to thermal
fluctuations. At $T=T_c$ both spins  are equally populated and the
bands becomes symmetric.
As expected we have a separated impurity band from main band for a
strong coupling $J=2.0t$ shown in  Fig. \ref{fig_1}(b).
When $n=x$ the impurity band is fully occupied, no low energy
hopping processes to main band are allowed and the system becomes a
band insulator. If the impurity band is partially occupied the
delocalization energy increases. At the half filling of the impurity 
band the system has the highest $T_c$ \cite{Chattopadhyay01a}.
As the temperature increases spin disorder grows and the band becomes
symmetric. But the impurity band width of the paramagnetic state
($T\ge T_c$) is smaller than that of the ferromagnetic state.
This band shrinking occurs because the
neighboring spins in the paramagnetic state are uncorrelated.

\begin{figure}
\includegraphics[width=2.8in]{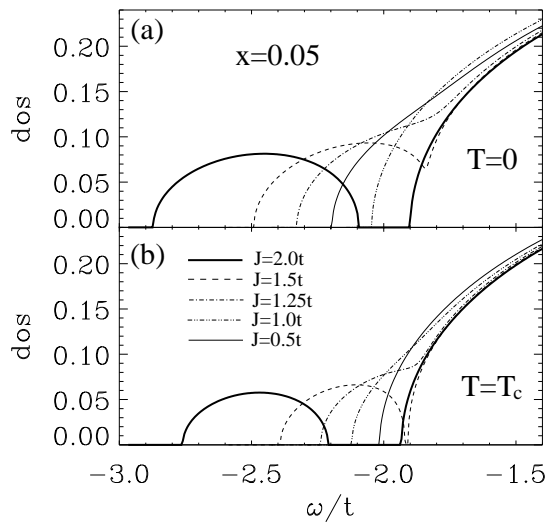}
\caption{
\label{fig_3}
The calculated majority spin DOS at (a) $T=0$ and (b) $T=T_c$
is shown for a fixed $x=0.05$ and for various coupling constant
$J/t=$ 0.5, 1.0, 1.25, 1.5, 2.0.
}
\end{figure}


In Fig. \ref{fig_2} we show the calculated majority spin DOS
corresponding to the
disordered spin state (at $T=T_c$, bottom panels) and
ferromagnetic state (at $T=0$, top panels).  The evolutions of the energy
($\omega$) dependent 
DOS are shown for  different doping parameter  $x=$0.001, 0.01, 0.05,
0.1 and for fixed 
coupling constant (a) $J=1.0t$ and (b) $J=1.5t$.
In our model the impurity level (acceptor energy level) and the formation
of an impurity band depend on the ferromagnetic coupling $J$. 
If $J \le J_c =t$ the impurity level  
is not isolated from the main band,
but if $J>J_c$ we find an isolated impurity level below the main band.
The small dot in Fig. \ref{fig_2}(b) indicates the isolated 
impurity level in the dilute limit ($x \rightarrow 0$). 
As $x$ increases for $J>J_c$ an impurity band  
centered around the impurity level is formed below the main band. 
For $J\le J_c$ the impurities give rise to band
tailing in the main band edge instead of forming impurity band. The
band width of the impurity band for $J>J_c$ increases with $x$
since the number of states of the impurity band increases with $x$.
If $x$ is bigger than $x_c$ the impurity band merges into the main band.
For $J=1.5t$ we have $x_c=0.032$ at $T=0$K and
$x_c = 0.071$ at $T=T_c$.

In Fig. \ref{fig_3} we show 
the majority spin DOS at (a) $T=0$ and (b) $T=T_c$
for a fixed $x=0.05$ and for various coupling constant
$J/t=$ $0.5$, $1.0$, $1.25$, 1.5t, 2.0t.
For $J\le J_c$ we see the expected
band shift proportional to $xJ$. For $J>J_{c}$ an impurity band centered
at impurity level and containing $x$ states is seen to split off from the
main band. Due to the widening of the impurity band with $x$ we find
the separated impurity band when $J>1.58t$ at 
$T=0$ and $J>1.45t$ at $T=T_c$ for $x=0.05$.
Thus the separated impurity band is
expected with small $x$ and large values of $J$.

\begin{figure}
\includegraphics[width=2.8in]{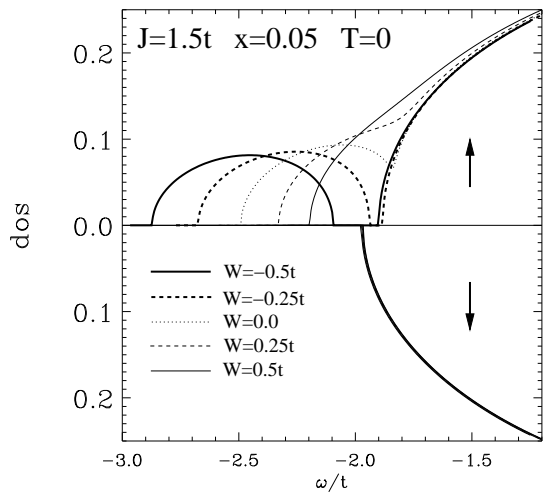}
\caption{
\label{fig_4}
The majority (top panel) and minority (bottom panel) DOS at $T=0$ 
are shown for a fixed value of $J=1.5t$ and $x=0.05$ for various 
potential scattering
$W/t=-0.5$, $-0.25$, 0.0, 0.25, 0.5.
}
\end{figure}

In Fig. \ref{fig_4} we show 
the calculated DOS at $T=0$ 
for a fixed value of $J=1.5t$ and $x=0.05$,
and for various values of potential scattering
$W=$$-0.5t$, $-0.25t$, 0.0, $0.25$, $0.5$.
The DOS of majority (minority) spin is shown in the top (bottom) panel.  
When we include the potential scattering ($W$) in addition to the
spin-spin coupling ($J$),
the carriers with spin parallel to 
local impurity spin feel a potential $-J+W$ on each magnetic impurity
site and anti-parallel carriers feel a potential $J+W$.
Thus the formation of the impurity band and the corresponding physical
properties of the system depend on the combined 
coupling $W\pm J$.
Fig. \ref{fig_4} shows that while the band edge of the
minority spins is
slightly dependent on the potential scattering, the majority spin DOS 
is strongly affected by the potential scattering. 
Even weak potential scattering can change the extended 
majority spin band into the 
spin-polarized impurity bands or the well formed impurity band into the
band tail of the main band.

In the following section we show that our calculated resistivities 
depend strongly on whether the carriers are within the impurity band
or in the band tail of the extended main band.


\section{dc-resistivity}

The conductivity is calculated from the usual Kubo formula. 
The Kubo formula for the conductivity $\sigma$ involves the two
particle current-current response function. 
Since the irreducible vertex in the response function is purely local 
in our approximation
of DMFT there is no vertex correction \cite{Pruschke}.
Thus, within this approximation only the simple bubble survives and
the real part of the finite frequency
conductivity is given by
\begin{eqnarray}
\sigma(\Omega,T) & = & e^2 \sum_{\sigma} \int d\varepsilon 
D(\varepsilon)\Phi(\varepsilon)
\int \frac{d\omega }{\pi }\frac{\left[ f(\omega )-f(\omega +\Omega
)\right] }{\Omega }\nonumber \\
&\times&  A_{\sigma} (\varepsilon ,\omega) A_{\sigma}
(\varepsilon ,\omega +\Omega),
\label{sigxx}
\end{eqnarray}
where $A_{\sigma}(\varepsilon ,\omega) = -(1/\pi) {\rm Im}
G_{\sigma}(\varepsilon ,\omega)$ is the spectral function,
$\Phi(\varepsilon)
=(4t^2-\varepsilon^2)/3$ is the current vertex for the Bethe lattice
\cite{Chattopadhyay00}, $f(\omega)$ is the Fermi distribution
function. For the hypercubic lattice \cite{Pruschke}
$\Phi(\varepsilon)=1$ has been 
used in Eq. (\ref{sigxx}), but for the Bethe lattice the explicit form
is derived in ref. \onlinecite{Chattopadhyay00}.
The dc-resistivity $\rho=1/\sigma_{dc}$ is then found from
Eq.~(\ref{sigxx}) in 
the limit $\Omega \rightarrow 0$ with $\sigma_{dc} = \sigma(\Omega
\rightarrow 0)$, which is given by
\begin{equation}
\sigma_{dc}(T) = e^2 \sum_{\sigma} \int d\varepsilon 
D(\varepsilon)\Phi(\varepsilon)
\int \frac{d\omega }{\pi }\left ( -\frac{\partial f(\omega)}{\partial
    \omega} \right ) A_{\sigma}^2(\varepsilon,\omega).
\end{equation}


\begin{figure}
\includegraphics[width=3.0in]{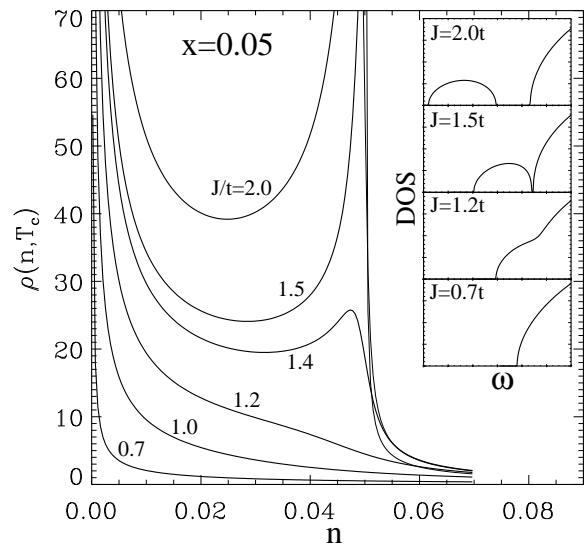}
\caption{
\label{fig_5}
The density dependent resistivity at $T=T_c$
for a fixed value of $x=0.05$ and for
various  coupling constant
$J/t=0.7$, 1.0, 1.2, 1.4, 1.5, 2.0. Insets show the DOS with the
corresponding parameters. 
}
\end{figure}

In Fig. \ref{fig_5} we show the calculated resistivity at $T=T_c$
as a function of carrier density 
for a fixed value of $x=0.05$ and various coupling constants $J/t$=0.7,
1.0, 1.2, 1.4, 1.5 and 2.0. Insets show the
density of states near  the band edge corresponding to the
disordered spin state ($T=T_c$).
All calculated results show that $\rho$ diverges as $n\rightarrow 0$
due to the absence of carriers.
As density increases we find two different behaviors depending on
the formation of the impurity band. (The critical coupling constant
which gives rise to the formation of the well separated impurity band
below the main band for $x=0.05$, $T=T_c$ is $J_c=1.48t$.)  
In the weak coupling limit $J \le 1.0 t$, where the impurity band
formation dose not happen,  the resistivity decreases as
the density increases monotonically. However, in the strong coupling
limit $(J \ge J_c)$, where the impurity band is formed, the
resistivity diverges again when the impurity bands are fully filled
(i.e., for $n=x$) and the system becomes a band insulator. In
the intermediate coupling regime ($1.0t < J < J_c$) the resistivity
shows non-monotonic behavior. 


\begin{figure}
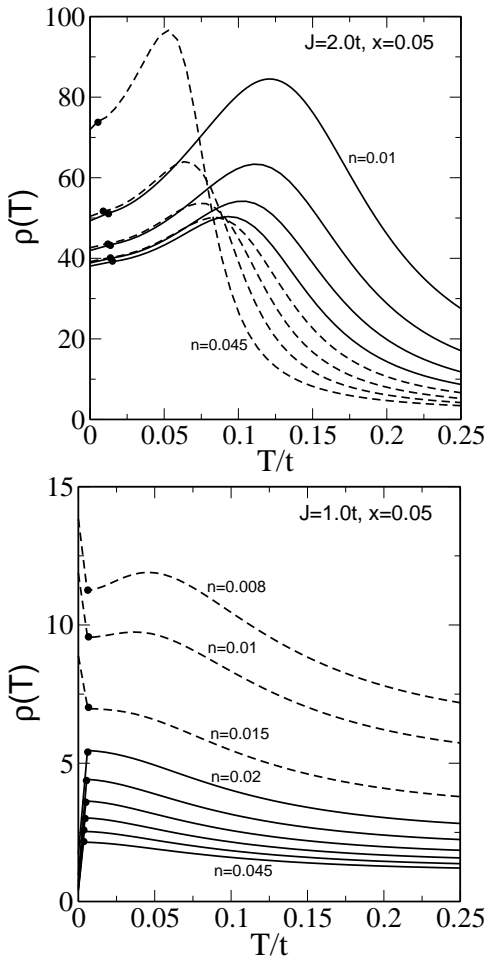

\includegraphics[width=2.5in]{fig_6a.eps}
\includegraphics[width=2.5in]{fig_6b.eps}
\caption{
\label{fig_6}
The calculated resistivity for (a) a  strong coupling
$J=2.0t$ with densities $n=$ 0.01, 0.015, 0.02, 0.025, 0.03
0.035, 0.04, 0.045 (top to bottom)and (b) a weak coupling $J=2.0t$
with densities 
$n=$ 0.008, 0.01, 
0.015, 0.02, 0.025, 0.035, 0.04, 0.045 (top to bottom). 
Dots indicate $T_c$ for given densities.  
}
\end{figure}

\begin{figure}
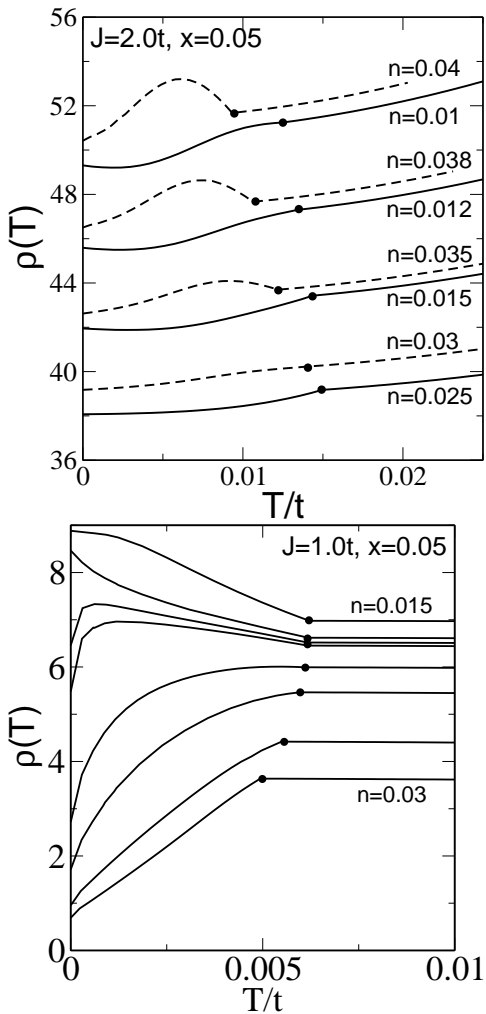

\includegraphics[width=2.50in]{fig_7a.eps}
\includegraphics[width=2.50in]{fig_7b.eps}
\caption{
\label{fig_7}
The low temperature dc-resistivity as a function of temperature 
for a fixed $x=0.05$ and for various densities. In (a) the results for a
strong coupling constant ($J=2.0t$) and densities
$n=0.01$, 0.012, 0.015, 0.025, 0.3, 0.35, 0.38, 0.40 are shown. 
In (b) we use $J=1.0t$ and
$n=0.015$, 0.016, 0.0163, 0.0165, 0.18, 0.2, 0.25,
0.3 (from top to bottom). Dots indicate the 
$T_c$ for given parameters.
}
\end{figure}

In Fig. \ref{fig_6} we 
show the calculated resistivity as a function of temperature for
various density.
For the strong coupling limit ($J=2.0t$), in Fig. \ref{fig_6}(a),
we find a crossover (resistivity peak) separating a good metal at low
$T$ from a semiconductor  at higher $T$.
The resistivity peaks are proportional to the energy separation
between the chemical 
potential and the band edge of the minority band.
In the high temperature regime above the resistivity peak the decreasing
resistance with increasing temperature, characteristic of a
semiconductor or insulator, is due to thermal excitation of the
carriers from impurity
band to the upper minority spin band. The
resistivity decreases with density because the carriers in the main
band are scattering off the impurities. But at low temperatures
most carriers in the impurity band contribute to the scattering and
give non-monotonic density dependence of the resistivity.
The metallic behavior at low temperature can be understood by the
disappearance of the coherent central quasiparticle peak in the DOS. 
For the weak coupling limit ($J=1.0t$), in Fig. \ref{fig_6}(b),
the resistivity can be explained by the scattering effects in the main
band except the behavior in the ferromagnetic state
($T<T_c$). For $T<T_c$ the resistivity changes from insulating at low
densities to metallic at high density.
Details of this behavior are given in Fig. \ref{fig_7}. 
We see that the resistivity crossover takes place only at low
densities since the minority band is
occupied by the carriers at high density.
 
In Fig. \ref{fig_7} we show the dc-resistivity as a function of
temperature for different densities in the low temperature regime ($T<2T_c$). 
In Fig. \ref{fig_7}(a) we use the parameters $x=0.05$ and a strong
coupling $J=2.0t$.  In this case all carriers (if $n \le x$)
occupy  the impurity band and stay mainly at Mn sites. Thus, the
carriers in the impurity band follow the fluctuation of the localized
Mn spin. 
At half filling of the spin polarized impurity
band ($n=0.025$) the resistivity has the lowest value 
and its behavior corresponds to that of the
double exchange (DE) model (see Fig. \ref{fig_10}), that is, the
resistivity decreases below the critical 
temperature because of  the spin disorder scattering as the
temperature decreases \cite{furukawa}.    
In the low density limit ($n<0.25$, less than half filling) the 
resistivity is dominated by the `impurity band' contribution and the
resistivity increases as density decreases 
due to the lack of mobile carriers.
As the carrier density is increased above the half filling ($n>0.25$) 
the resistivity increases 
due to the filling of the band.
We also find a very different temperature behavior of the resistivity
from the low density case.
As the temperature decreases the resistivity increases just below $T_c$,
then decreases at very low temperatures. 
The counter-intuitive increase of resistivity just below $T_c$ 
(as $T$ is decreased) arises because, as the carrier spins are aligned to
the impurity spins, the binding of the carriers to the 
impurity spins increases
corresponding to an increase in the basic scattering rate. 
In the very low temperature range, however, the DE-like mechanism 
dominates, which gives rise to decrease of the resistivity.
(When the impurity band is spin-polarized, carriers
which are bound to impurity 
site must have spins parallel to impurity spin. Thus, 
as the spins order ferromagnetically, the
basic ability of carriers to move 
in the impurity band increases.) The overall 
temperature dependence of resistivity 
is very weak in the strong coupling limit.
This weak $T$-dependence of the dc resistivity below $T_c$ occurs 
because the increase in scattering rate due to the binding is compensated 
by DE-like mechanism.

In Fig. \ref{fig_7}(b) we show the results for a weak coupling
limit, $J=1.0t$. In this case the impurities contribute to form the
band tail of 
the main band and the chemical potential lies in the main band. 
In the low density limit ($n<n_c=0.0164$, where $n_c$ is the density
above which the minority band starts to fill)
the resistivity below $T_c$ increases as the temperature decreases.
The increase in resistivity
is due to increased carrier-spin coupling as mentioned above. 
But, in the high density limit ($n>n_c$, where at 
$T=0$ the minority spin-band is occupied) 
we find the resistivity decreasing as the temperature decreases.
This metallic behavior in high density regime and for a low coupling
constant can be understood by the small scattering rate of the
carriers in the minority band.
If the carriers are near the edge of the majority spin band,  the
carriers form a spin-polarized bound state, so 
the effective scattering rate strongly increases, which gives rise to
the insulating behavior as shown in the low density results.
On the other hand, if the carriers are in the minority spin band the
carriers form 
an anti-bound state at the top of the 
band, so that at the physically relevant lower band edge, 
the effective scattering rate decreases. 
In addition in three dimensions the vanishing of the density of states 
at the band edge further decreases the scattering. 
These effects are quite large and dominate as the
minority band is occupied, which gives rise to the metallic behavior
in the high density limits. 
In the paramagnetic state the resistivity is almost temperature
independent. In the limit $T>T_c$ the local Green function $G(\omega)$ is
temperature independent and the resistivity depends on temperature only
weakly through the Fermi function (thermal smearing around the
chemical potential). 

\begin{figure}
\includegraphics[width=3.0in]{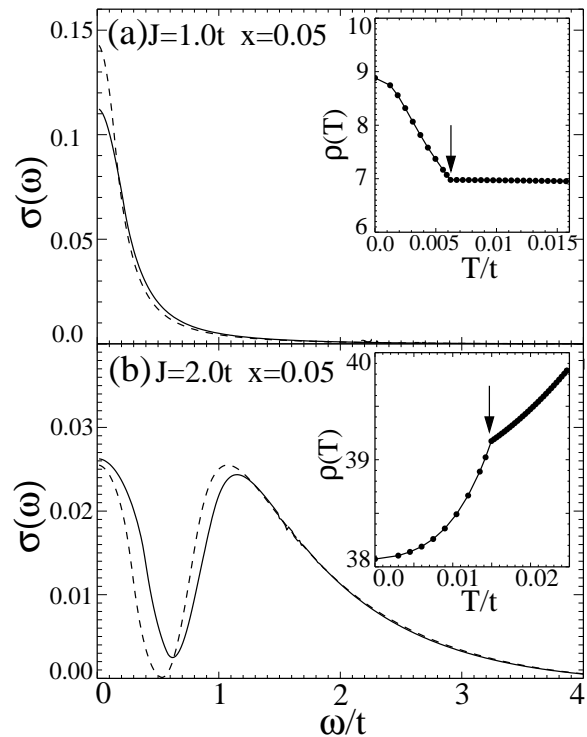}
\caption{
\label{fig_8}
The energy evolution of optical conductivity for (a) a intermediate
coupling 
$J=1.0t$ and (b) a strong coupling $J=2.0t$. Solid (dashed) lines
correspond to the results for $T=0$ ($T=T_c$). Insets show the dc
conductivities with corresponding parameters. Arrows indicate $T_c$.  
}
\end{figure}

In Fig. \ref{fig_8} we show the relation between the optical
conductivity and dc-resistivity.
The main panels of Fig. \ref{fig_8} show the evolution of the
conductivity for two couplings; weak ($J=1.0t$,
Fig. \ref{fig_8}(a), where the impurity band formation is not
accomplished),  and 
strong ($J=2.0t$, Fig. \ref{fig_8}(b), where the impurity band is
well formed); 
the insets show the dc resistivity with the same parameters.
The solid (dashed) lines indicate the optical conductivity for $T=0$
($T=T_c$).
When the impurity band is not formed
(for $J=1.0t$) we find approximately the Drude form for 
optical conductivity expected for carriers  
scattering off random impurities (a closer examination reveals minor 
differences due to density of states variations near the band edge). 
In this case (and for $n<n_c$) the dc conductivity shows insulating
behavior due to the  
formation of the bound state. Since the carrier density is low enough
not to fill the minority band, the formation of
anti-bound state dose not take place, which reduce the scattering rate.
In the $J=2.0t$ case the density of states plot shows the formation of 
an impurity  
band and the corresponding conductivity has two structures: 
a low-frequency quasi-Drude peak corresponding to motion within the 
impurity band and a higher frequency peak corresponding to excitations 
from the impurity band to the main band. In this case the dc
resistivity shows 
metallic behavior because the reduction of spin-disorder scattering
dominates over the bound-state formation. In In$_{1-x}$Mn$_x$As
\cite{hirakawa} we find the Drude-like conductivity is
correlated with the insulating behavior, but in Ga$_{1-x}$Mn$_x$As
\cite{singley}  the mid infrared peak in the optical conductivity is
closely related to the decrease of the dc-resistivity below the
critical temperature.

\begin{figure}
\includegraphics[width=3.0in]{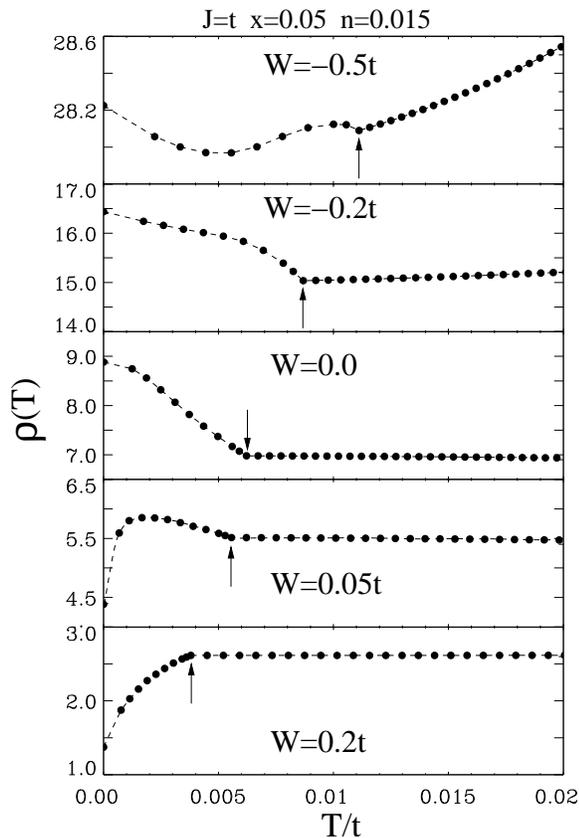}
\caption{
\label{fig_9}
The calculated dc-resistivity as a function of temperature
for a fixed $x=0.05$, $n=0.01$, $J=1.0t$, and for
various  potential scattering
$W/t=-0.5$, $-0.2$, 0.0, 0.05, 0.2. Arrows indicate the 
critical temperature $T_c$.
}
\end{figure}

Fig. \ref{fig_9} 
shows the sensitivity of the predicted behavior to potential 
scattering. In this figure we use the parameters: $x=0.05$, $n=0.01$,
$J=1.0t$, and for various  potential scatterings
$W/t=$$-0.5$, $-0.2$, 0.0, 0.05, 0.2.
At zero scalar potential ($W=0.0$, middle panel) the
impurity band  is not formed, and the resistivity shows insulating
behavior below $T_c$ due to the bound state formation of the carrier spins
with impurity  spins.
As the potential is made more attractive (negative),
the impurity  band features become  pronounced and the spin-disorder
scattering decreases (the carriers  move easily 
in the impurity band).
As the potential is made more repulsive (positive), 
the impurity band rapidly rejoins the main band. This reduces the
energy gap between the Fermi energy and the band edge of the minority
band, and for large enough repulsive potential the minority band
starts to be filled by the carriers. Thus, the anti-bound state 
formation dominates and it reduces the scattering rate below $T_c$ and
shows a metallic behavior. 
The DMS transport properties are sensitive to the combined 
coupling $W\pm J$,  and not solely to the exchange coupling $J$.

\begin{figure}
\includegraphics[width=2.2in]{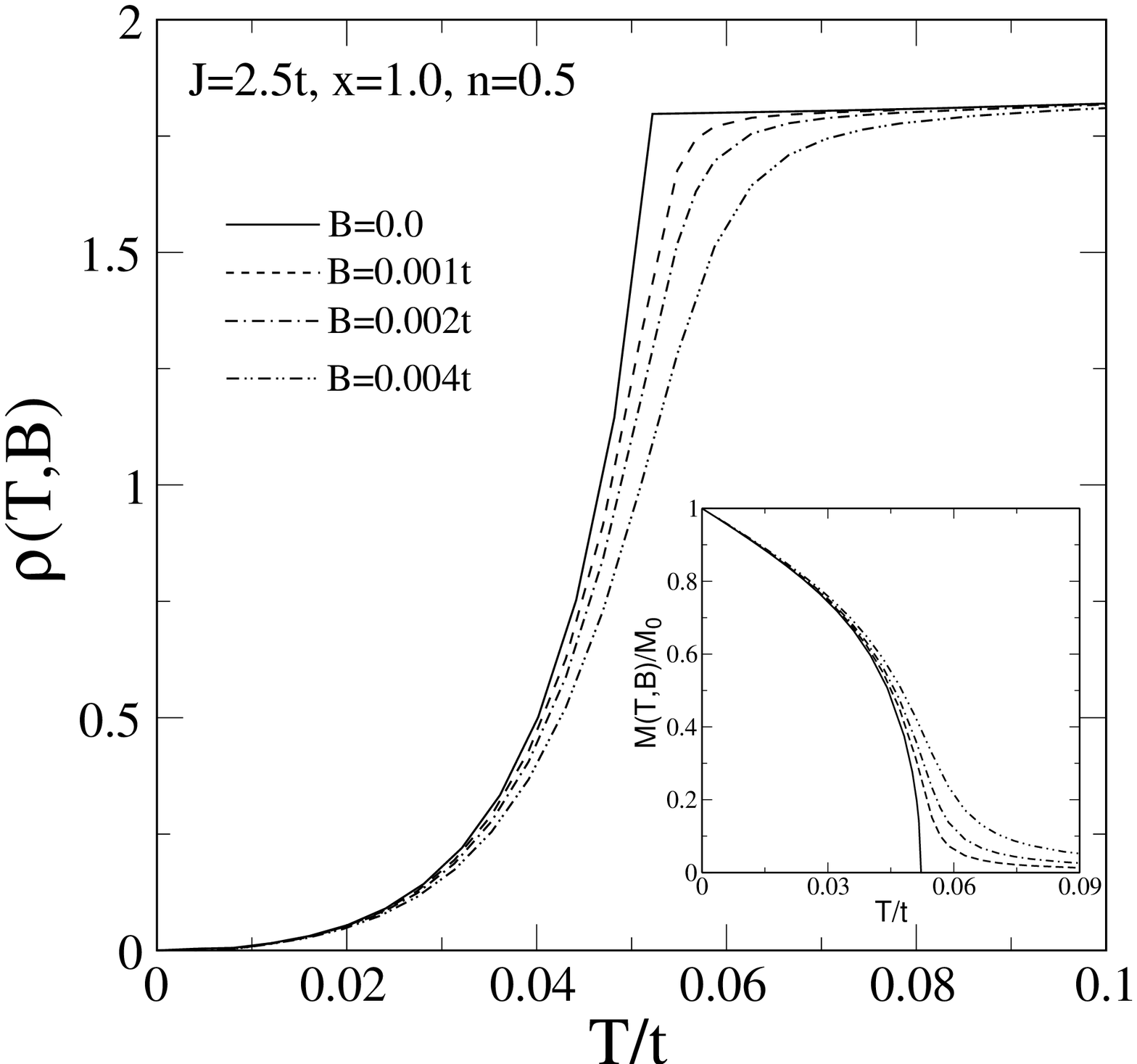}
\includegraphics[width=2.2in]{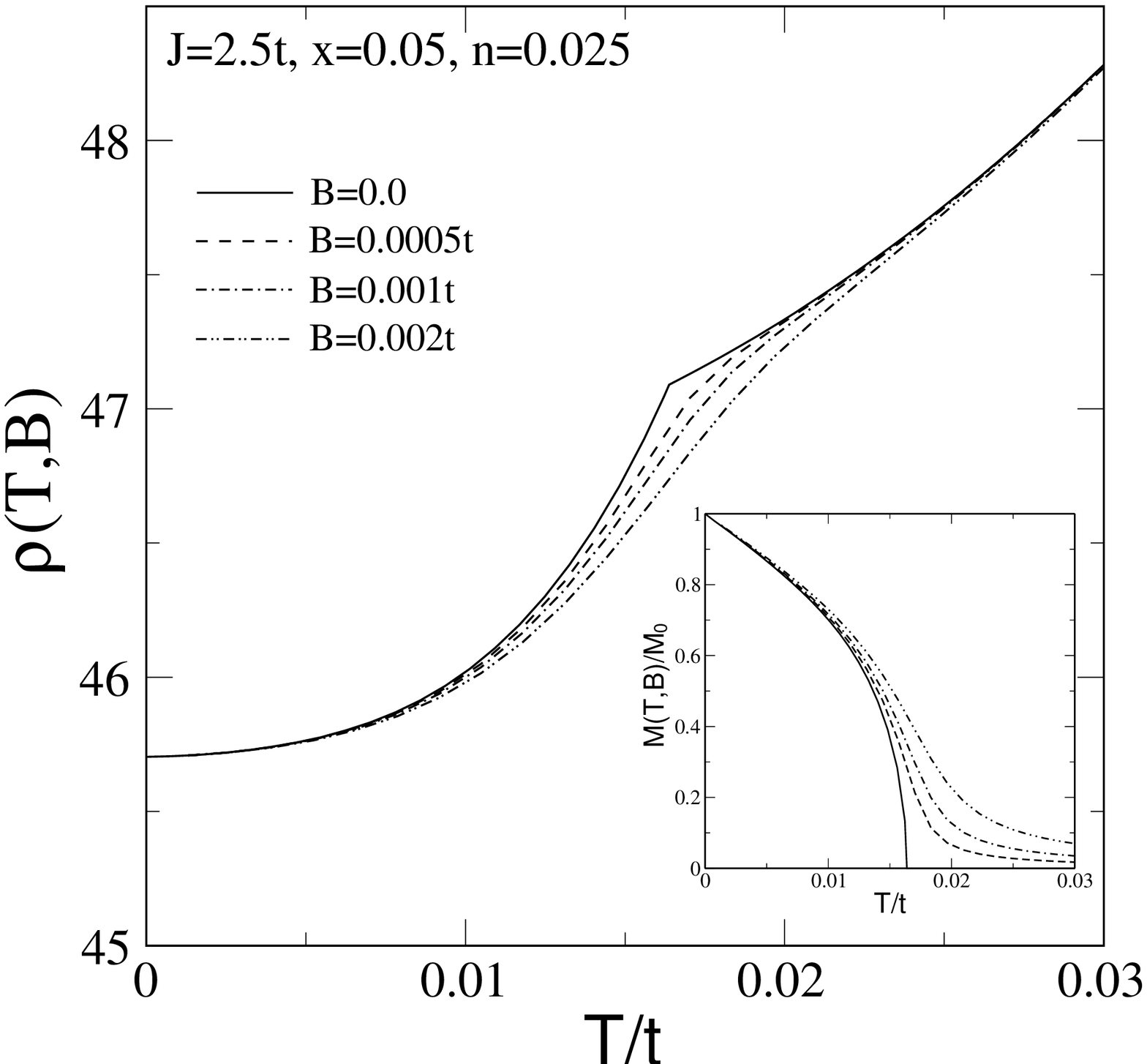}
\caption{
\label{fig_10}
Resistivity $\rho(T,B)$ for various magnetic field is given as a function of
temperature for (a) $J=2.5t$, $x=1.0$, $n=0.5$, and (b)
$J=2.5t$, $x=0.05$, $n=0.025$. Insets show the normalized
magnetization $M/M_0$, where $M_0$ is the saturation magnetization.
}
\end{figure}

Now we compare the transport properties of DMS
to those of another system with strong carrier-spin 
couplings, namely CMR manganites \cite{CMRReview}. In CMR,
instead of being dilute random impurities as in DMS,
the Mn ions form an ordered lattice. They 
possess large local moment, to which mobile carriers are very 
strongly coupled. Thus instead of a spin-polarized impurity band, 
there is a spin-polarized conduction band,
sufficiently 
well separated from other spin bands.
The periodic arrangement of the Mn sites means that (in the
absence of other physics) the scattering rate decreases as T is lowered.
CMR materials can be understood well by the
double exchange (DE) model \cite{furukawa}. In our model this corresponds to 
$x=1$, in which all magnetic ions
replace Ga at the cation sites. In this case ($x=1$) 
the fully polarized spin band is well separated from the other bands
instead of forming impurity band.
The temperature dependences of the resistivity $\rho(T,B)$ and
magnetization $M(T,B)$ are given in Fig. \ref{fig_10}(a) for 
$x=1.0$ with $J=2.5t$, $n=0.5$, and various magnetic fields.
Above $T_c$ the resistivity has a small temperature dependence
since the local spin fluctuation is
saturated above $T_c$. Below
$T_c$ resistivity decrease as magnetization increases. The origin of
the resistivity dependence on the magnetization is spin
disorder scattering, which 
gives rise to the scaled behavior of the resistivity,
$\rho(M)/\rho(M=0) = 1-CM^2$, where $C$ is a temperature/field
independent constant \cite{furukawa}.
The origin of the resistivity is qualitatively explained by the
carrier scattering due to the thermally fluctuating spin
configurations, or the spin-disorder scattering. As the spontaneous or
the induced magnetic moment is developed, the amplitude of the spin
fluctuation decreases so that the resistivity also decreases. 
In Fig. \ref{fig_10}(b) we show the resistivity for DMS system
with $x=0.05$. The overall features of temperature and magnetic field
dependence look similar to the DE model. However, we find that the
negative magnetoresistence at $T_c$ is very weak and the resistivity
above $T_c$ is not saturated. The fast drop of the resistivity just
below $T_c$ can be explained by spin-disorder scattering.
These ideas of DE model have limited applicability to the DMS systems, i.e.
only for strong couplings and near half filling of the spin polarized
impurity band. As shown in previous figures
these ideas cannot explain  the resistivity behavior in low coupling
and high density regimes. 
Note that an important ingredient of DMS transport properties is
missing from our DMFT theory which, while accounting well for the
non-perturbative effects of spin disorder ($J$) and local potential
($W$) scattering by the magnetic impurities, leaves out all ionized
impurity disorder that may very well be important.


\section{electrical resistivity: Boltzmann transport approach}

When the dominant scattering mechanism is the scattering by charged
impurities the Boltzmann transport theory may be used to calculate the
electrical resistivity of the carriers since DMFT is not well-suited
for treating long-range disorder.
Due to the band splitting in the ferromagnetic state the
carrier densities $n_{\pm}$ for spin up/down are not equal. Note that 
the total density $n=n_+ + n_-$. In this
situation the total conductivity can be expressed as a sum of spin
up/down contributions
\begin{equation}
\sigma =\sigma_+ + \sigma_-,
\end{equation}
where $\sigma_{\pm}$ is the conductivity of the ($\pm$) spin subband.
The conductivities $\sigma_{\pm}$ are given by
\begin{equation}
\sigma_{\pm} = \frac{n_{\pm}e^2 \langle \tau_{\pm}\rangle }{m},
\end{equation}
where $m$ is the carrier effective mass, and the energy averaged
transport relaxation time $\langle \tau_{\pm} \rangle$ for the ($\pm$)
subbands are given in the Boltzmann theory by
\begin{equation}
\langle \tau_{\pm} \rangle  = \frac{\int d\varepsilon
  \tau_{\pm}(\varepsilon) \varepsilon \left [ -\frac{\partial
  f^{\pm}(\varepsilon)}{\partial \varepsilon}  \right ]} 
  {\int d\varepsilon \varepsilon \left [ 
    -\frac{\partial f^{\pm}(\varepsilon)}{\partial \varepsilon} \right
  ]},
\label{tau}
\end{equation}
where $\tau_{\pm}(\varepsilon)$ is the energy dependent relaxation
time for the ($\pm$) subbands, and $f^{\pm}(\varepsilon)$ is 
the carrier (Fermi) distribution function 
\begin{equation}
f^{\pm}(\varepsilon) =\frac{1}
{1+e^{[\varepsilon-\mu^{\pm}(T)]/k_BT}},
\end{equation}
where $\mu^{\pm}(T)$ is the chemical potential at finite temperature.
The energy dependent relaxation time is given in 
the Born approximation by
\begin{equation}
\left[\tau(\varepsilon_{\bf k})\right ]^{-1} = \frac{2\pi}{\hbar}
\int \frac{d^3k'}{(2\pi)^3}N_i|V^{sc}_{\bf k - k'}|^2 (1-\cos\theta_{\bf kk'})
\delta(\varepsilon_{\bf k} - \varepsilon_{\bf k'}),
\end{equation}
where $N_i$ is the charged impurity concentration, and  $V^{sc}_{\bf k -
  k'}$ is the screened carrier-impurity 
Coulomb interaction, which can be expressed as
\begin{equation}
V^{sc}(q) = \frac{4\pi Ze^2}{\kappa q^2}\frac{1}{1+[q_s(q)/q]^2},
\end{equation}
where $Z$ is the charge of impurities, $\kappa$ the background
(GaAs) lattice dielectric constant, and $q_s$ the 
temperature dependent screening function.

\begin{figure}
\includegraphics[width=2.2in]{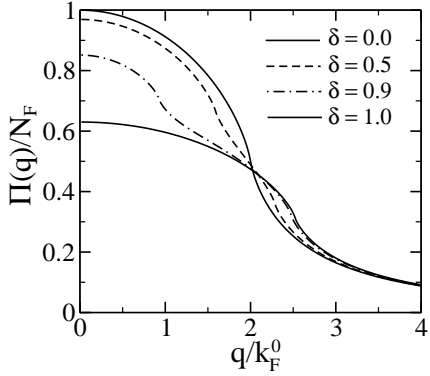}
\caption{
\label{fig_11}
Wave vector dependence of the effective screening function
(polarizability function $\Pi(q)$ normalized its paramagnetic state
value $N_F = \Pi^{\rm para}(0) = mk_F^0/\pi^2$) for various
polarizations $\delta =
(n_+ - n_-)/n$.
}
\end{figure}

In this paper we consider two screening approaches: Thomas-Fermi (TF)
approximation and random phase approximation (RPA).
Within TF screening we have 
\begin{equation}
q_s^2 = \frac{q_{TF}^2}{2}\left [ \frac{k_F^+}{k_F^0} +
  \frac{k_F^-}{k_F^0} \right ],
\end{equation}
where $q_{TF}^2 =
4 k_F^0/\pi a_B$ with $k_F^0 = (3 \pi^2 n)^{1/3}$ being the Fermi
wave vector of the spin 
unpolarized state and $a_B = \kappa
\hbar^2/me^2$  the effective Bohr radius. $k_F^{\pm} = (6 \pi^2
n_{\pm})^{1/3}$ is the Fermi wave vector of the each spin-split subband.
Note that TF screening wave vector, being a long wavelength
approximation, is dependent only on the spin polarization, but not on the
temperature explicitly. It is therefore
temperature independent above the critical temperature.
Within TF screening approximation we have the energy dependent relaxation
time (by integrating Eq. 21)
\begin{equation}
\left[\tau(\varepsilon_{\bf k})\right ]^{-1} =
\frac{2\pi}{3}\frac{N_i}{n}\frac{E_{F}^0}{\hbar} \frac{k k_F^0}
{\tilde{k}_F^2} F(4k^2/q_{s}^2), 
\end{equation}
where $\tilde{k}_F = (k_F^+ + k_F^-)/2$ and $F(x) = (2/x^2) [ \ln(1+x)
- x/(1+x)]$.  
When a system is fully polarized we have $k_F^+ = 2^{1/3}k_F^0$ and
$q_s = q_{TF}/2^{1/3}$.  
For the random phase approximation (RPA) the temperature
dependent screen function is given by
\begin{equation}
q_s^2(q,T) = \frac{q_{TF}^2}{2} \left [ \frac{k_F^+}{k_F^0}\Pi_+(q,T) +
  \frac{k_F^-}{k_F^0}\Pi_-(q,T) \right ], 
\end{equation}
where $\Pi_{\pm}(q,T)$ is the temperature dependent static
Lindhard function for each spin subband. 
In Fig. \ref{fig_11} we show the screening function
$q_s^2(q,T=0)/q_{TF}^2 = 
\Pi(q)/N_F$, where $N_F = \Pi^{\rm para}(0) = mk_F^0/\pi^2$, 
for various polarization $\delta =
(n_+ - n_-)/n$. For an unpolarized state ($\delta = 0$) there is an
inflection point at $q=2k_F^0$, which is the usual Kohn anomaly. For
partially polarized state 
$0<\delta<1$ we have two inflection points at $q = 2k_f^{\pm} =
(2n^{\pm}/n)^{1/3}$ and the value of $\Pi(q=0)/N_F$ decreases as
$\delta$ increases. When the system is fully polarized $\delta=1$, and 
$\Pi(q)$ has an inflection point at $q=2k_F^+$, and 
$\Pi(q=0) = N_F/2^{2/3}$.

\begin{figure}
\includegraphics[width=2.5in]{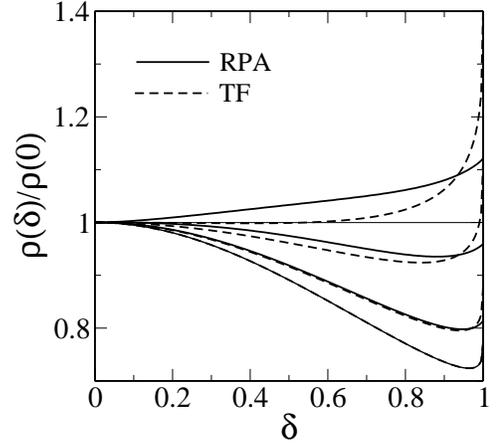}
\caption{
\label{fig_12}
Calculated resistivity as a function of $\delta = (n_+ - n_-)/n$ for
various value 
of $x_0 =$ 1.0, 2.0, 5.0, 10.0 (top to bottom).
Solid (dashed) lines show the results calculated with RPA (TF)
screening function.}
\end{figure}

At $T=0$, $f^{\pm}(\varepsilon) \equiv
\theta (E^{\pm}_F-\varepsilon)$ where $E^{\pm}_F$ is the Fermi energy
for the ($\pm$) subbands, and then $\langle
\tau_{\pm} \rangle \equiv \tau(E^{\pm}_F)$ giving the familiar result $\sigma
\equiv \rho^{-1} = n_{+}e^2\tau(E^+_F)/m + n_{-}e^2\tau(E^-_F)/m$. 
Within TF screening the ratio of the resistivity of the fully
polarized state to that 
of the unpolarized state becomes $\rho(\delta=1)/\rho(\delta=0) =
2^{5/3} F(2^{4/3}x_0)/F(x_0)$, where $x_0 = (2k_F^0/q_{TF})^2$.
(Note $x_0 \propto n^{1/3}$.)
In Fig. \ref{fig_12} we  show the calculated resistivity
as a function of $\delta = (n_+ - n_-)/n$ for
several value of $x_0$. Solid (dashed) lines show the results
calculated with RPA (TF) screening. 
In general, for small values of $x_0$, $\rho(\delta=1) > \rho(\delta
=0)$, but for large $x_0$, $\rho(\delta=1) < \rho(\delta=0)$. At $x_0
\approx 3$ we have $\rho(\delta=1) = \rho(\delta=0)$. For GaAs $x_0
=3$ corresponds to the hole density $n \approx 10^{19} cm^{-3}$.
In relatively high density limits (large $x_0$)
the two approximations agree very well, which indicates that the $q=0$
scattering  mostly contributes to the scattering time.
However, in the low density regime (small $x_0$) we that find the two
screening theories give very different results for the 
spin polarized state. Noting that the TF approximation is just the
long wavelength limit of RPA, we emphasize that RPA is obviously
the better theory. 

\begin{figure}
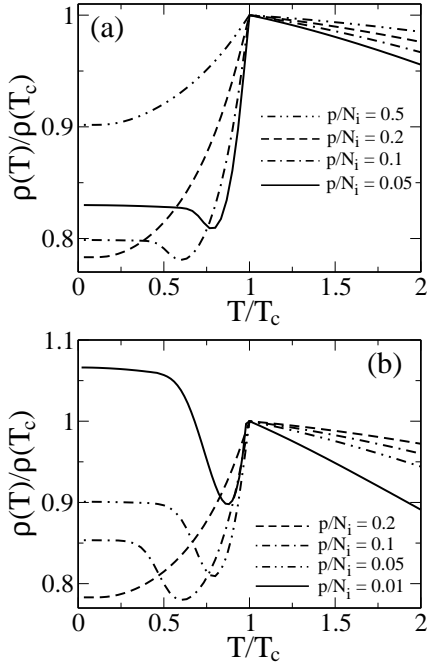

\includegraphics[width=2.2in]{fig_13a.eps}
\includegraphics[width=2.2in]{fig_13b.eps}
\caption{
\label{fig_13}
Calculated resistivity, $\rho(T)/\rho(T_c)$,
as a function of temperature. In (a) the RPA results are shown with
hole densities $p/N_i =$ 0.05, 0.1, 0.2, 0.5. In (b) the TF screening
results are shown with hole densities $p/N_i =$ 0.01, 0.05, 0.1, 0.2.
Here the impurity density $N_i=10^{21} cm^{-3}$ is used.
}
\end{figure}

\begin{figure}
\includegraphics[width=2.2in]{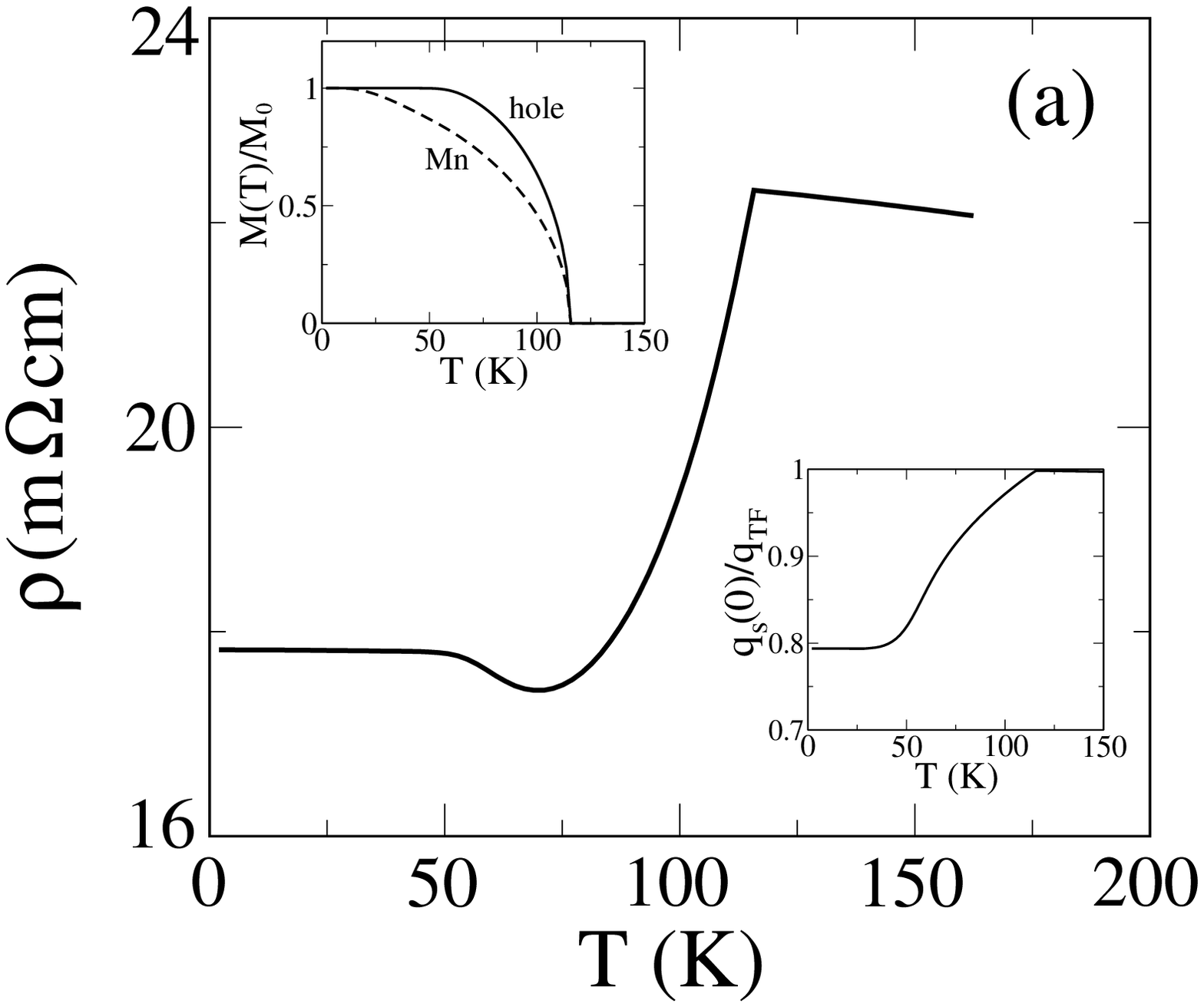}
\includegraphics[width=2.2in]{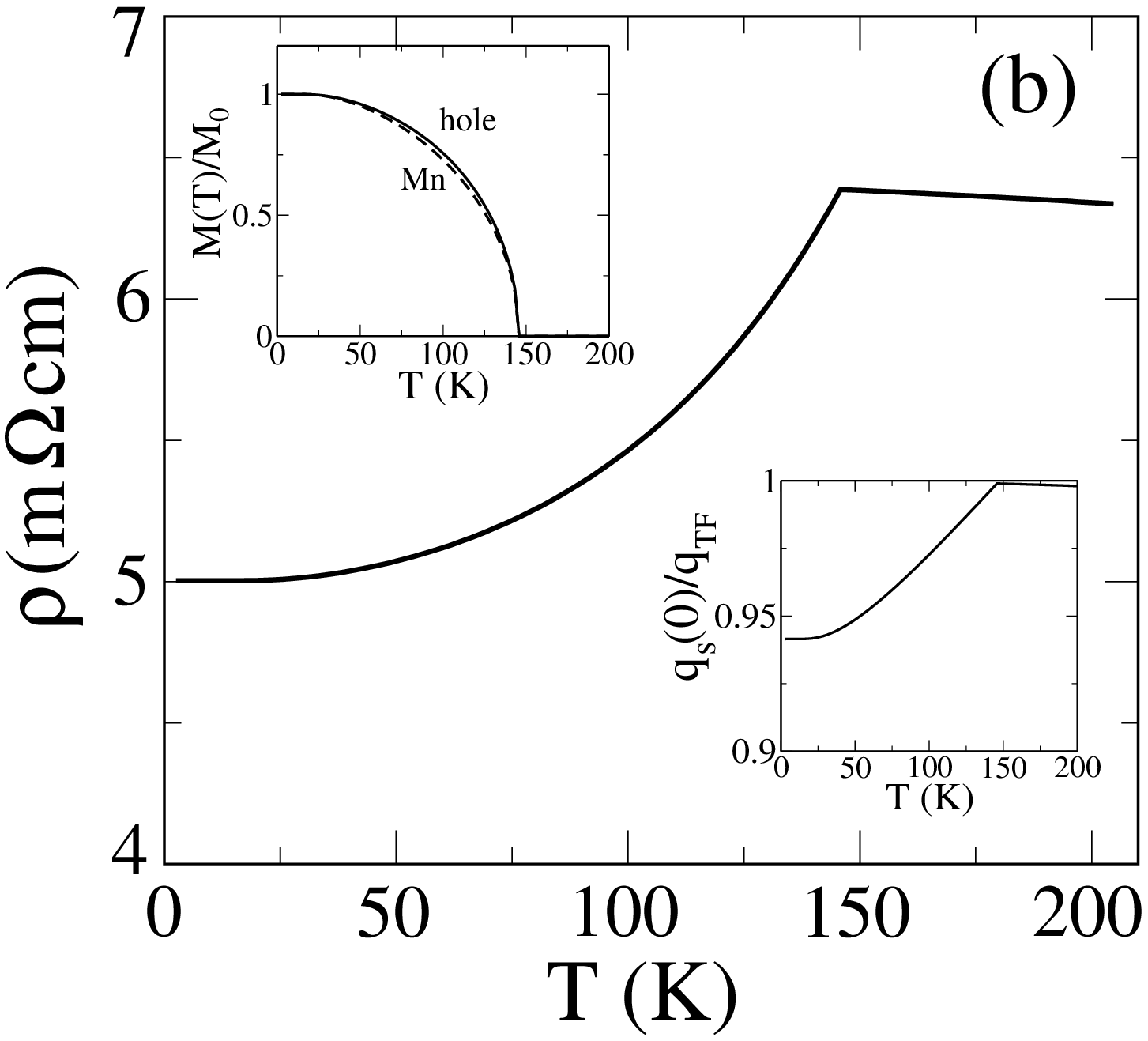}
\caption{
\label{fig_14}
Calculated resistivity
as a function of temperature for
a hole density (a) $p/n_i =$ 0.1 and (b) $p/N_i=$0.2 with finite
temperature RPA screening function.
The top insets show the temperature dependence of the Mn ions and hole
magnetization. The bottom insets show the calculated temperature
dependence of the TF screening wave vector $q_s(0)$. 
Here the impurity density $N_i=10^{21} cm^{-3}$ is used.
}
\end{figure}

In Figs. \ref{fig_13} and \ref{fig_14} we show our calculated
resistivity for GaMnAs samples. We use the parameters
corresponding to  GaMnAs: dielectric constant $\kappa = 12.9$,
hole effective mass $m=0.5m_e$, impurity density $N_i=10^{21}cm^{-3}$,
and the magnetic coupling constant $J=$120 meV nm$^3$.
Fig. \ref{fig_13} shows normalized resistivity
$\rho(T)/\rho(T_c)$ as a function of 
temperature $T/T_c$ for several values of hole density. 
In Fig. \ref{fig_13} there are two independent sources of temperature 
dependence in our calculated resistivity --- one source is the energy 
averaging defined in Eq. (\ref{tau}) and the other is the explicit
temperature dependence of the finite  temperature RPA screening
function $q_s(q,T)$. Since the Fermi temperature is much higher than
the critical temperature 
($T_c/T_F \ll 1$) the screening function is weakly temperature
dependent above the critical temperature (unpolarized state). Thus
the decrease of the resistivity above $T_c$ with increasing
temperature arises from the thermal energy averaging. This is a
well-known high-T effect of ionized impurity scattering in
semiconductors: $\rho(T)$ decreases with increasing $T$ due to the
thermal averaging. As density decreases this effect is enhanced. 
In the ferromagnetic polarized state ($T<T_c$), as the temperature
decreases from $T=T_c$, the screening length increases until all the
holes are polarized. In this temperature range we find a strong
temperature dependent resistivity.
This strong temperature dependence arises from the 
low temperature screening function and the change of the carrier
densities of the each subband.  
When all the carriers are polarized (i.e. at low density and
temperature) the screening function
is almost independent of temperature and the resistivity is also  
therefore temperature independent.
The interplay between the screening length and the down spin (--)
carrier density gives rise to the minimum of the resistivity in the
low density regime. The decrease of the scattering time due to the fast
increase of the down spin density overwhelms the increase of the
scattering time due to the decrease of the
screening length, producing the non-monotonic behavior for $p/N_i \le
0.1$ and $T/T_c \sim 0.5$ in Fig. 13.
But in the high density regime, where the spins are not fully
polarized even at low $T$, screening is the dominant effect on
the temperature dependent resistivity.
As the density increases the relative low temperature resistivity,
$\rho(T)/\rho(T_c)$, decreases until the holes are partially
polarized. The change of screening wave vector is larger in this
case, leading to the monotonically increasing $\rho(T)$ in the $T\le
T_c$ regime.

In Fig. \ref{fig_14} we show our calculated resistivity
as a function of temperature for
two hole densities (a) $p/N_i =$ 0.1 and (b) $p/N_i=$0.2 with finite
temperature RPA screening function.
The temperature dependence of the impurity (i.e. Mn) moment
magnetization as well as the hole spin polarization is 
given in the insets, which is calculated using 
the Weiss molecular mean-field theory for delocalized carriers
\cite{dassarma03}. Note that the hole gas is almost fully spin
polarized upto 
$T/T_c = 0.5$ at low density, but at higher density the holes are
partially polarized even at $T=0$.
The bottom insets show the calculated temperature
dependence of the finite temperature screening wave vector
$q_s(0)$. At high density 
the screening wave vector changes by 6\% when the temperature
increases from zero to $T_c$ due to the partial polarization of the
holes at $T=0$. But at low density the decrease of the screening wave
vector is about 20\%.  
In the metallic GaMnAs samples the change of the resistivity when
the temperature goes from $T_c$ to zero is about 20\% in good
agreement with our calculation. 
Similarly the observed decrease of $\rho(T)$ for $T>T_c$ also arises
naturally in our theory as a consequence of thermal averaging.
Thus, the temperature dependence
of the resistivity in the metallic DMS samples may be arising almost
entirely from the temperature dependence of screening and thermal
averaging in the charged impurity scattering.

\section{discussion}

We have developed in this work two complementary theories for
understanding DMS transport properties. Our work establishes that DMS
transport, even in ideal intrinsic circumstances, is rather
complex, and depends sensitively on many system parameters including
the exchange coupling, the magnetic impurity density, the carrier
density, the temperature, the band parameters of the parent
semiconductor (e.g. effective mass, etc.), and the details of the
charged impurity distribution (and therefore compensation). Given such
a complex parameter space, it is not meaningful to try to develop
quantitative transport theories at this early stage of the subject
since all the intrinsic parameters may not be known. We therefore
focus in this work on developing a comprehensive qualitative
theoretical description which emphasizes general broad features of how
various parameters affect DMS transport behavior. As such we have
concentrated in this work on understanding temperature and carrier
density dependence of dc resistivity in a model DMS system, keeping
primarily the extensively studied Ga$_{1-x}$Mn$_x$As system in
mind. Even for GaMnAs, the transport data for various values of Mn
concentration ($x \approx 0.01-0.1$) cover much too broad a range of
behavior for a unified and coherent theoretical description. For
example, low (and sometimes large) Mn concentrations ($x \le 0.03$ and
sometimes $x >0.05$) are known to lead to insulating transport
behavior usually attributed to a disorder-driven metal-insulator
(Anderson) localization. We neglect {\it all} localization effects in
our theory. The localized GaMnAs regime in all likelihood requires its
own characteristic theoretical description such as the bound magnetic
polaron percolation theory \cite{kaminski1} whose transport properties
\cite{kaminski2} have recently been theoretically analyzed.

Even without the disorder-induced strong localization complications,
neglected completely in this work, we face the formidable difficulty
of using the semiconductor valence/conduction band (the valence band
for GaMnAs, where the carriers mediating the ferromagnetic interaction
are holes) or the magnetic impurity induced impurity band (i.e. Mn
induced d-band in the fundamental band gap of GaAs) picture for
describing the carrier dynamics. The precise nature (i.e. valence
band versus impurity band) of the DMS carriers is still a
controversial issue although it is likely that at the very high doping
densities (e.g. Mn density $\sim 10^{21}cm^{-3}$ in GaMnAs) of DMS
interest the impurity band overlaps strongly with the valence band
(i.e. forms the tail of the valence band), and therefore, the
distinction between the valence and the impurity band picture is not
a real qualitative difference. Our DMFT theory, presented in sections
II-IV of this paper, clearly shows that in the strong exchange
coupling ($J/t \gg 1$) regime the impurity band picture applies
whereas in the weak-coupling regime ($J/t <1$), the valence band
picture applies. It is possible that GaMnAs belongs to the
intermediate coupling regime ($J/t \sim 1$), where it may be more
appropriate to think of the holes to be residing in the extended tail
of the valence band, presumably with an enhanced effective mass
compared with the GaAs valence band hole mass. Such a coupled
impurity-valence band picture of GaMnAs is consistent with recent
optical spectroscopy measurements \cite{hirakawa,singley}, but more
experimental work is needed to settle this question.

The theoretical strength of our DMFT description is that, being a
nonperturbative technique, it can handle both strong-coupling and
weak-coupling regimes, and our results presented in Figs. 6 -- 10 of
this paper show qualitative difference between the strong-coupling
regime ($J/t=2$) with an impurity band well-separated from the
semiconductor band and the intermediate-coupling regime ($J/t=1$) with
only band-tailing and no separate impurity band
formation. Temperature, carrier density, and impurity concentration
all play qualitatively important roles in determining the dc transport
properties within DMFT, and sorting out the details with respect to
experimental results may be extremely difficult.

The weakness of DMFT is that it can only include spin-disorder
scattering (controlled by $J$) and Mn impurity induced local potential
scattering (controlled by $W$) effects on transport properties. As
such it leaves out the most important scattering mechanism which may
be operational in real samples, namely scattering by charged
impurities which is often the most important resistive scattering
process in heavily doped semiconductos below the room temperature (or
the optical phonon scattering regime which may well be above the room
temperature). The reason DMFT is unable to account for charged
impurity scattering is that the Coulombic impurity potential is long
ranged, and DMFT by construction is a local theory. Thus, rewriting
our starting Hamiltonian (Eq. 1) more completely we have
\begin{equation}
H=H_{host} + H_M + H_{AF} + [H_i + H_c],
\end{equation}
where $H_i$ is the carrier-charged impurity interaction and $H_c$ is
the carrier-carrier (i.e. hole-hole in GaMnAs) interaction. In
principle, the terms (i.e. $H_i$, $H_c$) within the square bracket are
parts of the $H_{host}$, but it is important to appreciate their
considerable (perhaps even dominant) importance in determining the dc
transport properties. To include the charged impurity scattering
effects on transport, we use the highly successful and robust
semiclassical Boltzmann transport theory to DMS systems assuming a
mean-field approach where the long-range Coulomb impurity potential
arising from $H_i$ (assuming random impurity scattering) is screened
by the polarization bubble diagrams arising from $H_c$. This type of
Boltzmann transport theory is extremely successful in describing
semiconductor transport properties \cite{dassarma}. We note that our
DMFT and Boltzmann transport theories are complementary -- DMFT treats
the spin disorder and the local potential scattering associated with
Mn dopants and Boltzmann theory treats the scattering by screened
charged impurity scattering.

The magnetic DMS properties enter the Boltzmann theory only indirectly
through the carrier spin polarization calculations. Spin disorder
scattering is {\it not} explicitly included in the Boltzmann theory
although it is straightforward to do so. Our Boltzmann theory
manifests nontrivial interplay among temperature dependent screening,
temperature dependent spin polarization (i.e. spin up-down carrier
densities), and thermal energy averaging, leading to temperature
dependent resistivity (Figs. 13 and 14) that are rather similar to
experimental observations \cite{Matsukura98} in GaMnAs. Based on this
qualitative similarity we conclude that much of GaMnAs DMS transport
is dominated by screened charged impurity scattering with spin
disorder scattering playing only a rather minor quantitative
role. Recently, Lopez-Sancho and Brey \cite{brey2} have come to a very
similar conclusion for GaMnAs transport properties.

\begin{figure}
\includegraphics[width=2.2in]{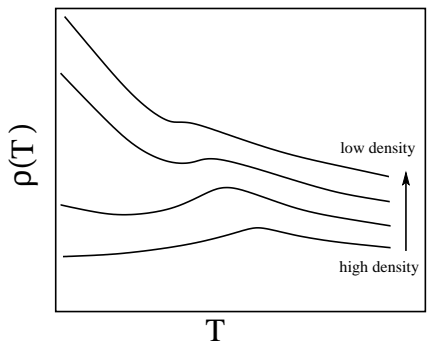}
\caption{
\label{fig15}
Schematic diagram for the experimental GaMnAs dc-resistivity for
decreasing hole density (from bottom to top) (see ref. [3]).
}
\end{figure}

Before concluding we now discuss our theoretical results in light of
the existing DMS transport data. Although there are some experimental
transport data in other DMS systems (most notably InMnAs with
qualitatively similar behavior to GaMnAs), truly extensive and
reliable transport data\cite{Matsukura98} are available only for
Ga$_{1-x}$Mn$_x$As (in the $x\approx 0.01-0.08$ regime) DMS
samples. Even for this well-studied system, experimental transport
results are problematic, and have considerable spread in the sense
that nominally ``identical'' GaMnAs samples (i.e. same nominal carrier
density and Mn concentration made in the same growth run) may have
different $T_c$ and transport properties. This situation is improving
as sample quality and processing (e.g. annealing) techniques improve,
but experimental DMS transport properties are still not robust in a
quantitative way. This is of course very understandable given the very
large parameter space (i.e. exchange coupling, hole density, Mn
concentration, defects and impurities, compensation, band structure
parameters, etc) that DMS transport depends on. With these serious
caveats we show in Fig. 15 a schematic depiction of the generic
experimental observation for $\rho(T)$ in GaMnAs for various hole
densities. At high hole density the system shows ``metallic'' behavior
for $T<T_c$ with $\rho(T)$ increasing somewhat with $T$ upto $T_c$,
and then decreasing slowly for $T>T_c$. Even the optimally doped most
``metallic'' GaMnAs is, however, at best a bad metal with mobilities
of the order of 10 $cm^2/Vs$ with $k_Fl \sim 1$ where $l$ is the
transport mean free path. It is important to realize that DMS
transport is always extremely highly resistive due to the very large
amount of impurities and defects invariably present in the  low
temperature MBE process needed for producing homogeneous GaMnAs
samples. Thus, from the perspective of heavily doped semiconductors,
although the DMS systems may be above the Mott limit (i.e. the carrier
density in the Mott metallic range), they are close to being Anderson
insulators due to strong disorder effects. As hole density decreases
the system eventually becomes an insulator at low enough carrier
densities (the top curve in Fig. 15), with $\rho(T)$ decreasing
monotonically increasing $T$.

Two generic features of Fig. 15 are: (a) a peak (or a kink) in
$\rho(T)$ near $T \approx T_c$, and (2) the slow decrease of $\rho(T)$
for $T>T_c$. Both of these generic features of $\rho(T)$ as well as
the ``metallic'' high-density behavior are qualitatively (perhaps even 
semi-quantitatively) well-explained by our Boltzmann theory approach
including only scattering by ionized impurity scattering. This is
apparent by comparing Fig. 15 with Figs. 14 and 13 where our Boltzmann
theory results are shown. Physically, the increasing $\rho(T)$ with
$T$ ($<T_c$) arises from the decreasing strength of screening due to
the interplay of two independent and competing physical effects:
Temperature induced suppression of screening and spin polarization
(i.e. the spin polarization decreasing with increasing temperature)
induced enhancement of screening with increasing temperature. As
explained in Sec. V the competition between these two effects depend
on the carrier density, leading to some weak non-monotonicity in
$\rho(T)$ for $T<T_c$. In this screening picture, the resistivity peak
or cusp at $T\approx T_c$ arises from the ferromagnetic to
paramagnetic transition which affects screening as the carrier density
of states (which is inversely proportional to the screening length)
changes from one in the fully spin polarized state to two in the fully
paramagnetic phase. Thus, even without any spin disorder scattering 
effects, just screened ionized impurity scattering by itself will give
rise to the peak or the kink in $\rho(T)$ at $T=T_c$. We believe that
spin disorder scattering, which also produces a kink at $T\approx T_c$
(see, for example, Figs. 7 and 9), play only a minor role in the
resistivity ``maximum'' at $T_c$ in optimally metallic GaMnAs with
most of the peak structure arising from the screening properties of
ionized impurity scattering. The second generic experimental feature
in Fig. 15, the slow decrease of $\rho(T)$ for $T>T_c$, cannot be
explained {\it at all} by spin disorder scattering since spin disorder
should remain large in the paramagnetic system ($T\ge T_c$) and
certainly should not decrease with increasing $T$. Our Boltzmann
theory provides a natural explanation (Figs. 13 and 14) for the
decreasing $\rho(T>T_c)$ as arising from the energy averaging of the
relaxation time (Eq. 19), i.e., $\rho(T)$ decreases with increasing
$T$ simply because the holes move ``faster'' at higher temperatures
(i.e. increasing kinetic energy with increasing $T$). This decreasing
$\rho(T)$ with increasing $T$ ($>T_c$) also shows that our neglect of
phonon scattering in the transport theory is a valid approximation
since phonon effects will always increase $\rho(T)$ with increasing
$T$. $\rho(T)$ will increase again  when phonon
scattering starts dominating over ionized impurity scattering at much
higher temperatures. The relative lack of importance of phonon
scattering in DMS systems arises from their very strong charged
impurity resistive scattering effects as reflected in very small
sample mobilities.


\section{conclusion}

In this paper we investigate the transport properties of the diluted
magnetic semiconductors using dynamical mean field theory and
Boltzmann transport theory.
We have shown that the resistivity depends strongly on the
system parameters, i.e., exchange coupling, carrier density, doping,
and temperature. The resistivity drop with decreasing temperature in
the ferromagnetic state can be 
partially explained by the screening theory for metallic samples.
The parameter dependence of the resistivity 
contains important information about the 
physics of diluted magnetic semiconductors.
We find that in the strong exchange coupling regime the spin disorder
scattering and the formation of the bound state in the impurity band
compete to produce an 
unusual behavior in the temperature dependent resistivity.
We also show that in the weak coupling regime
the occupation of the minority spin band is critical to the scattering
mechanisms, and substantially reduces the resistivity 
because the repulsive interaction between local moments and 
``wrong-spin'' carriers suppresses the carrier amplitude at the 
impurity site, reducing the effective carrier-spin coupling.
Our Boltzmann transport theory for charged impurity scattering is good
qualitative agreement with the existing DMS experimental data, showing
that transport in DMS GaMnAs system may very well be dominated
primarily by screened ionized impurity scattering (with spin disorder
scattering playing only a minor secondary role), at least for the
optimally doped metallic GaMnAs samples. We have completely neglected
detailed band structure complications (e.g. spin-orbit coupling in the
valence band) in our theory. These effects are certainly very
important, but our interest in this paper is the development of a
conceptually coherent qualitative theory for DMS transport identifying
the main transport mechanisms, and as such we have neglected all
nonessential complications.

\section*{ACKNOWLEDGEMENT}

We thank DARPA and US-ONR for support.


\end{document}